\documentclass[fleqn,usenatbib]{mnras}

\usepackage{newtxtext,newtxmath}

\usepackage[T1]{fontenc}

\DeclareRobustCommand{\VAN}[3]{#2}
\let\VANthebibliography\thebibliography
\def\thebibliography{\DeclareRobustCommand{\VAN}[3]{##3}\VANthebibliography}

\usepackage{graphicx}	
\usepackage{amsmath}	

\usepackage{amssymb,gensymb}	
\usepackage{floatrow,sidecap,multicol,booktabs,longtable}
\usepackage{caption}
\usepackage{makecell}
\usepackage{pifont}
\newcommand{\angstrom}{\text{\normalfont\AA}}
\usepackage[usenames,dvipsnames,table]{xcolor}
\usepackage{xltabular}
\usepackage{multirow}
\usepackage{hhline}
\newcommand\MUV{$M_\textrm{UV}$}
\newcommand{\spitzer}{\textit{Spitzer}}

\title[SuperBoRG: UV luminosity function]{The UV luminosity functions of Bright $z>8$ Galaxies: Determination from $\sim0.41$ deg$^2$ of HST Observations along $\sim 300$ independent sightlines}

\author[N. Leethochawalit et al.]
{
Nicha Leethochawalit,$^{1,2,3}$\thanks{E-mail:nicha.leethochawalit@unimelb.edu.au}
Guido Roberts-Borsani,$^{4}$
Takahiro Morishita,$^{5}$
Michele Trenti,$^{2,3}$
\newauthor
and Tommaso Treu,$^{4}$
\\
$^{1}$National Astronomical Research Institute of Thailand (NARIT), Mae Rim, Chiang Mai, 50180, Thailand\\
$^{2}$School of Physics, Tin Alley, University of Melbourne, VIC 3010, Australia\\
$^{3}$ ARC Centre of Excellence for All Sky Astrophysics in 3 Dimensions (ASTRO 3D), Australia \\
$^{4}$Department of Physics and Astronomy, UCLA, 430 Portola Plaza, Los Angeles, CA 90095-1547, USA\\
$^{5}$IPAC, California Institute of Technology, MC 314-6, 1200 E. California Boulevard, Pasadena, CA 91125, USA
}

\date{Accepted XXX. Received YYY; in original form ZZZ}

\pubyear{2015}

\begin{document}
\label{firstpage}
\pagerange{\pageref{firstpage}--\pageref{lastpage}}
\maketitle

\begin{abstract}
We determine the bright end of the rest-frame UV luminosity function (UVLF) at $z=8-10$ by selecting bright $z\gtrsim 8$ photometric candidates from the largest systematic compilation of HST (pure-)parallel observations to date, the Super-Brightest-of-Reionizing-Galaxies (SuperBoRG) data set. The data set includes $\sim300$ independent sightlines from WFC3 observations, totalling $800-1300$ arcmin$^2$ (depending on redshift). We identify 31 $z\gtrsim8$ candidates via colour selection and photo-$z$ analysis with observed magnitude ($24.1< H_{160} <26.6$) and $1\sigma$ range probability of $58-95\%$ of being true high-z galaxies. Following detailed completeness and source recovery simulations, as well as modelling of interloper contamination, we derive rest-frame UVLFs at $z=8-10$ down to $M_{UV}\simeq-23$. We find that the bright end of the galaxy luminosity function can be described both by a Schechter and by a double power-law function, with our space-based large area determination showing some tentative discrepancies with the luminosity functions derived from ground-based observations at the same redshifts. We identify two possible explanations. The first is that the UVLF includes a substantial contribution from AGNs at magnitudes $M_{UV}<-22$. The UVLF we derive at $z=8$ is consistent with no evolution of the bright end from $z=6-7$ UVLFs that include AGNs. An alternative (and non-exclusive) possibility is that the dust content of the observed galaxies decreases with redshift, thereby countering the naturally expected decrease in the UV luminosity functions because of a decreased star formation rate with increasing redshift. Both scenarios raise interesting prospects to further understand galaxy formation in extreme objects during the epoch of reionisation.   
\end{abstract}




\section{Introduction}
Galaxies in the first $\sim 700$ Myr from the Big Bang, well into the epoch of reionisation at redshift $z\gtrsim 8$, have been primarily observed in rest-frame ultraviolet (UV) light through space-based Near-Infrared (NIR) data (e.g., see \citealt{Bouwens2015})
Galaxies at this epoch are young and predominantly star-forming, given the limited time since the Big Bang and the rapid growth of dark matter halos and stellar mass assembly at those early times (e.g. \citealt{Mason2015}). With stellar populations comprising primarily of young stars, to the first order of magnitude, the galaxy spectrum is a power-law function. Interstellar and intergalactic neutral gas absorbs photons bluer than the Lyman $\alpha$ wavelength, introducing a sharp spectral break at $\lambda \sim 0.12 (1+z)~\mathrm{\mu m}$, which is at the base of the so-called dropout or Lyman-break colour selection \citep{Steidel1996}.  Therefore, rest-frame UV tends to be the brightest part of the spectra and incidentally also the most easily accessible for relatively large area observations that are deep enough for source identification, thanks to the Wide Field Camera 3 (WFC3) on the Hubble Space Telescope (HST). Note, however, that once galaxy candidates have been identified, follow-up observations are possible, with a large sample of objects having detections/upper limits on rest-frame optical fluxes from Spitzer IRAC \citep[e.g.,][]{Laporte2015, Roberts-Borsani2016, Stefanon2022, Finkelstein2022}, and far-IR from ALMA, including spectroscopic confirmations based on the [O III] line \citep[e.g.,][]{Inoue2016, Tamura2019, Laporte2021}.

One of the most fundamental quantities that can be extracted from the observations of rest-frame UV images is the number density of sources per unit volume, i.e. the UV luminosity function (UVLF) \citep[e.g, ][]{Steidel1999, Bouwens2007}. Many physical properties of the galaxies and halos can be inferred from UVLFs, such as the number of ionising photons \citep{Mason2015}, a cosmic star-formation rate density history \citep{Tacchella2018}, dust properties at high redshifts \citep{Ma2019}, and the connection between stellar mass and dark matter halo assembly \citep{Behroozi2019, Ren2019}. In the past decade, there has been substantial effort to constrain UVLFs at $z\gtrsim8$ and to investigate whether the shape can be well described by a Schechter function $\Phi(L) \propto (L/L_*)^{\alpha} \exp{(-L/L_*)}$ \citep[e.g.,][]{Schmidt2014, Bouwens2015, Calvi2016, Morishita2018, Stefanon2019, Bowler2020, Rojas-Ruiz2020, Bouwens2021}. Generally, studies based on deep imaging agree that the faint-end slope $\alpha$ continues to get steeper with increasing redshift \citep{Finkelstein2015, Ishigaki2018, Bouwens2021}, although determinations at \MUV$\gtrsim -15$ mag are based on small samples of strongly lensed sources, and will thus need deeper observations to be firmly established.

The bright end of $z\gtrsim 8$ UVLFs is still not well constrained. Bright galaxies are rare. A large survey area is necessary to obtain a statistical number of bright galaxies, especially in the presence of bias due to cosmic variance \citep{Trenti2008}. One method to reduce the impact of cosmic variance is to make observations in several independent pointings. The idea has led to several programs taking advantage of pure-parallel random-pointing opportunities with HST, in particular through the Brightest of Reionizing Galaxies (BoRG) \citep{Trenti2011, Bradley2012, Schmidt2014, Calvi2016, Morishita2021} and HIPPIES surveys \citep{Yan2011}. Alternatively, ground-based telescope surveys, such as COSMOS/UltraVISTA and VIDEO surveys \citep{Stefanon2019, Bowler2020}, used large coverage but mostly contiguous areas ($>1$ deg$^2$) to overcome the cosmic variance problem. At face values, the number densities derived by these surveys vary by up to 2 dex at \MUV\ brighter than $-22$ mag, albeit with large uncertainties of up to $\sim0.5$ dex due to Poisson noise. Generally, the UVLFs derived from HST pure-parallel surveys show higher number densities. For example, \citet{Calvi2016} found that with an inclusion of a particularly bright $M_\textrm{UV}\sim-22.75$ mag galaxy in the sample, their $z\sim9$ UVLF at $M_\textrm{UV}\sim-22$ bin shows an excess of $\gtrsim1$ dex at face value when compared to the predicted UVLF model by \citet{Mason2015}. A similar trend is found by \citet{Rojas-Ruiz2020} at $z=8$ in an independent re-analysis of the BoRG and HIPPIES datasets. In contrast, the studies that are based on ground-based observations \citep[e.g.,][]{Stefanon2019, Bowler2020} found number densities are generally lower than those derived with HST pure-parallel surveys. Ground-based luminosity function determinations are somewhat lower than theoretical expectations (e.g. \citealt{Mason2015} model) even though functionally they show an excess compared to the exponential drop-off component of the Schechter function (while the \citealt{Mason2015} model predicts a Schechter-like luminosity function at all redshifts). 
 
Both ground and space-based luminosity function determinations are challenging due to different systematic uncertainties and potential contamination. Objects at $z\gtrsim 8$ at the bright end are intrinsically very rare, opening up the possibility of substantial contamination from photometric scatter \citep{Vulcani2017}, particularly for pure-parallel datasets which lack legacy deep multi-wavelengths observations that are instead available for smaller areas. The data is also often inhomogeneous due to the nature of pure parallel opportunities. On the other hand, ground-based surveys usually contain observations from more photometric bands but at shallower depths. They are subjected to atmospheric absorption and poorer spatial resolution, which limit opportunities to discriminate between candidates and surrounding sources and between stars and compact galaxies \citep[e.g.,][]{Wilkins2014}. 
For example, \citet{Bowler2015} and \citet{Varadaraj2023} removed all candidates whose SEDs have $\chi^2<10$ when fit with stellar templates regardless of the $\chi^2$ when fit with the galaxy templates. This conservative removal of stars may cause some high-$z$ galaxies with similar colours to be excluded from the sample.

In this work, we aim to constrain the bright end of $z=8-10$ UVLFs based on the most comprehensive archival dataset of HST parallel fields, the SuperBoRG compilation \citep{Morishita2021}. SuperBoRG is the most extensive compilation of pure parallel and coordinated-parallel HST programs with WFC3 in the past decade. It consists of 316 sightlines with moderate-depth multi-band imaging data. Most of the data is part of the Brightest of the Re-ionizing Galaxies (BoRG) programs that have been conducted to date, totalling $\sim1400$ HST orbits (GO 11700, 12572, 12905, 13767, 14701, 15212, 16005, PI M. Trenti). Given a large number of sightlines, SuperBoRG is essentially free from cosmic variance. 

The structure of this paper is as follows. In Section \ref{sec:data_and_sample_selection}, we describe the sample selection (based on \citealt{Roberts-Borsani2021}) and UV magnitude derivation. In Section \ref{sec:completeness}, we describe the completeness simulation and effective volumes of the study. We derive the UV luminosity functions in Section \ref{sec:UVLF} and discuss the results in Section \ref{sec:discussion}. Throughout this paper, we quote magnitudes in AB system and use the \citet{Planck2016} cosmology, with $\Omega_m=0.315$, $\Omega_\lambda = 0.685$, $\Omega_b = 0.0490$, $h = 0.6731$, $\sigma_8= 0.829$, and $n_s = 0.9655$.

\section{Data and sample selection}
\label{sec:data_and_sample_selection}
This work is based on the SuperBoRG $z\sim8-10$ photometric candidates we derived and presented in \citet[][from here on RB22]{Roberts-Borsani2021}, where we derived their physical properties such as half-light radii, UV spectral slopes, and rest-frame optical properties. In this paper, we adopt the same sample and provide here a summary of both the data sets and the selection process used to identify the candidates (for full details, we refer the reader to Section 2 of RB22 and \citealt{Morishita2021}).

\subsection{Data}
The SuperBoRG project is a compilation of HST archival imaging data focused on parallel programs, which include BoRG \citep{Trenti2011}, the Hubble Infrared Pure Parallel Imaging Extragalactic Survey \citep[HIPPIES,][]{Yan2011, Yan2012}, and several random-pointing COS-GTO coordinated parallel observations with WFC3. It also includes two coordinated-parallel programs: CLASH \citep{Postman2012} and RELICS \citep{Coe2019}, which observed multiple parallel fields located at $\sim6$ arcmin away from massive galaxy clusters. \citet{Morishita2021} consistently processed all SuperBoRG images using the same in-house pipeline. All images are reduced, sky-subtracted, and PSF-matched using the F160W frame as a reference. The average $5\sigma$ depths for the 0.6''-diameter aperture are 26.7 mag for the F350LP filter, 26.6 mag for the F606W filter, 26.7 mag for the F098M filter, 26.6 mag for the F125W filter, 26.6 mag for the F140W filter, and 26.4 mag for the F160W filter.

The final number of fields included in our study depends on the redshift of the sample. To select candidates at a given redshift range, a field must include all filters required for a NIR dropout colour selection and at least one blue filter. Approximately $80\%$ of the fields were observed with at least 5 HST filters. Fields from cycle 18 of the HIPPIES program were excluded due to significant IR persistence in those images (see Section 2.1 of RB22). 

In addition to HST observations, the SuperBoRG dataset includes overlapping \textit{Spitzer}/IRAC 3.6 $\mu$m and 4.5 $\mu$m imaging, which covers approximately half of the data set. We summarise the number of fields and the total area used for sample selection at each redshift in Table \ref{tab:field_summary}. Depending on the redshift sample, the total area ranges from 840 to 1318 arcmin$^2$.

\begin{table}
\begin{tabular}{c|r|r|r|r|r|r|}
\cline{2-7}
                                      & \multicolumn{6}{c|}{ Survey area included in this analysis (arcmin$^2$)} \\ \hline
\multicolumn{1}{|c|}{z} & BoRG       & HIPPIES  & COS      & CLASH  & Relics   & Total      \\ \hline
\multicolumn{1}{|c|}{8}               & 826  & 42  & 91  & 13  & 81  & 1053 \\ 
\multicolumn{1}{|c|}{9}               & 712  & 23  &  0  & 13  & 91 & 840 \\ 
\multicolumn{1}{|c|}{10}              & 878  & 37  & 96  & 216 & 91 & 1318 \\ \hline
\end{tabular}
\caption{Total area used in this study to select galaxy candidates and determine UVLF at each redshift}\label{tab:field_summary}
\end{table}

\subsection{Sample selection}\label{subsec:sample_selection}
The $z=8-10$ Lyman break galaxies (LBGs) we presented in RB22 were selected as follows. Based on the SuperBoRG images, source catalogues were created with \texttt{SExtractor} \citep{BertinArnouts1996} in dual image mode. High-redshift galaxies are expected to have blue UV slopes and are thus likely to be most luminous in the band immediately redward of the Lyman break (or dropout filter). Therefore, the F125W filter was used as the detection image for the $z\sim8$ candidate selection and a stacked F140W+F160W image for the selection of $z\sim~9-10$ candidates. Photometric colours of each object were calculated using 0.6''-diameter aperture magnitudes. Total fluxes were derived by scaling each galaxy's 0.6''-diameter aperture flux by the ratio of its \texttt{AUTO} flux and its 0.6''-diameter aperture flux in a given reference band (F125W for the $z\sim8$ sample, and F140W+F160W for $z>8$ samples).

\subsubsection{Dropout selection and visual inspection}\label{sec:dropout_selection}
Dropout galaxies were subsequently selected from the source catalogues based on their photometric colours and standard colour selection criteria. The criteria adopted for $z\sim8$ galaxy candidates (Y-dropouts) in images where the $Y_{098}$ band was available were 
\begin{gather*}
    S/N_{J_{\rm125}} > 6.0 \,\,\land S/N_{H_{\rm 160}}>4.0 \,\,\land\\
    Y_{098}-J_{\rm125}>1.75 \,\,\land J_{\rm125}-H_{\rm 160}<0.5 \,\,\land\\ (J_{\rm125}-H_{\rm 160})<0.02+0.15\cdot(Y_{098}-J_{\rm125}-1.75),
\end{gather*}
and no detection (S/N$<$1.0) in the $I_{814}$ band (if present) and all other bluer filters. For fields without $Y_{098}$ coverage, but with $Y_{\rm 105}$ images instead, we used:
\begin{gather*}
    S/N_{J_{\rm125}} > 6.0 \,\,\land S/N_{H_{\rm 160}}>4.0 \,\,\land\\
    Y_{105}-J_{\rm125}>0.45 \,\,\land J_{\rm125}-H_{\rm 160}<0.5 \,\,\land\\ Y_{105}-J_{\rm125}>1.5\cdot(J_{\rm125}-H_{\rm 160})+0.45,
\end{gather*}
and also non-detection in the $I_{814}$ band (if present) and in all bluer filters. If both $Y_{\rm 098}$ and $Y_{\rm 105}$ imaging were available, an object was considered a high-$z$ candidate if it satisfied either one of the selection criteria.

For $z\sim9$ candidates ($JH_{\rm 140}$ dropouts), we required
\begin{gather*}
S/N_{JH_{\rm 140}}>6.0 \,\,\land
S/N_{H_{\rm 160}}>4.0 \,\,\land \\
Y_{105}-JH_{\rm 140}>1.5 \,\,\land 
JH_{\rm 140}-H_{\rm 160}<0.3 \,\,\land \\
Y_{105}-JH_{\rm 140}>5.33\cdot(JH_{\rm 140}-H_{\rm 160})+0.7,
\end{gather*}
and no detection in all filters bluer than $Y_{\rm 105}$. Lastly, the criteria for $z\sim10$ dropouts were
\begin{gather*}
S/N_{H_{\rm 160}}>6.0 \,\,\land \\
J_{\rm125}-H_{\rm 160}>1.3,
\end{gather*}
and no detection in $Y_{\rm 105}$ and all bluer filters. Given the primary source of contamination of high-$z$ samples are $z\sim2-4$ interlopers, we aimed to guard against such contamination by requiring all $z\sim8-10$ dropout candidates with \spitzer/IRAC 3.6 $\mu$m coverage not to display excessively red $H_{\rm 160}-[3.6]$ colours - i.e., we required $H_{\rm 160}-[3.6]<1.4$ (\citealt{Bouwens2015}). Additionally, we required that all candidates satisfy \texttt{SExtractor} stellarity parameter smaller than 0.95, mainly to avoid contamination from galactic cool stars (see, e.g. \citealt{Bradley2012}). 
Two authors (GRB and NL/TM) subsequently conducted a visual inspection of all the initial candidates satisfying the colour selection. The purpose of the inspection was to strengthen the sample selection by verifying two conditions. First, the non-detection in the filters blueward of the Lyman break must not be due to artefacts or lack of data. Second, the colours/fluxes in the IR bands must not be impacted by persistence, streak artefacts, hot pixels or speckles from nearby bright objects. Each inspector independently flagged each dropout candidate as `clear non-dropouts' or `suspicious' based on whether a candidate displayed any of these features and the degree to which they impacted the images/photometry.

Examples of suspicious candidates are those that showed optical flux in regions with high fluctuations due to surrounding noise, or those that reside near a particularly bright or extended object, making it more challenging to constrain the Lyman break and dropout: in such cases, objects have a high-$z$ nature which is somewhat unclear but not to the degree that warrants immediate exclusion. In the final catalogues, we discarded all candidates flagged as `clear non-dropouts' by either of the inspectors and those where both inspectors marked the object as 'suspicious'. The selection process yielded 108 candidates at $z=8$, 14 candidates at $z=9$, and 15 candidates at $z=10$.

\subsubsection{Photometric redshift selection refinement}
\label{sec:photometric_redshift}

The dropout candidates identified in RB22 were refined through a photometric redshift selection to reduce further the probability of having low-$z$ interlopers in the sample. Given the candidates were pre-selected with colour selections, the photo-$z$ analysis was conducted with an empirical prior created from simulated data and constructed over a grid of apparent magnitude and dropout redshift. Briefly, for each NIR selection, the distributions of the resulting galaxy redshifts in the simulated data sets were modelled with a double skew-normal profile and converted to a probability distribution: given the primary source of contamination resulting from such dropout criteria are $z\sim2-4$ interlopers, the double skew-normal profile provided an excellent fit to account for both the low-$z$ interlopers and high-$z$ selection without neglecting extended tails in the distributions but affording the flexibility for a Gaussian profile when needed. We refer the reader to Section 2.3.1 of RB22 for details on the construction of the prior and to Table 9 of the same paper for the best-fit parameters of the double skew-normal fits.

With the constructed priors in hand, the \texttt{EAzY} photometric redshift code \citep{Brammer2008} was then used to derive the redshift probability distributions of the candidates, adopting the default set of galaxy SED templates in version 1.3. To remain in the final sample, candidates must satisfy two conditions. Firstly, a given object was required to have a best-fit redshift ($z_{p}$) in the range $z_{p}=7.5-8.5$, $z_{p}=8.5-9.5$, or $z_{p}=9.5-10.5$ to be considered a $z\sim8$, $z\sim9$ or $z\sim10$ candidate, respectively. Secondly, the candidate was required to display a posterior redshift distribution such that $P(z>6.5)$ was greater than 0.7 for $z\sim8$ selection and $P(z>7.5)$ greater than 0.7 for $z\sim9-12$ selection. The selection process refined and reduced the initial dropout sample to 35 candidates at $z\sim8$, six at $z\sim9$, and two at $z\sim10$. 

\subsubsection{Removing and estimating potential contamination}
\label{sec:potential_contamination}
The final step in constructing the RB22 sample of $z\sim8$ galaxies was to exclude potential contaminations from spurious sources and brown dwarfs. Spurious sources originating from hot pixels can serve as potential contaminants, especially in pure-parallel observation where dithering is not feasible. To minimise the contamination, we used a strict combination of Source Extractor parameters, i.e. \texttt{DETECT\_THRESH }$=1$ sigma and \texttt{DETECT\_MINAREA} $=9$ pixels. \citet{Finkelstein2022} found that such a combination yields the least contamination when the sources are significantly brighter than the limiting magnitude of the detection images. Furthermore, we inspected the underlying RMS maps for all candidates to search for evidence of hot pixels. If a hot pixel was found in the immediate vicinity of the source, we rederived the photometry to exclude such pixels and reran \texttt{EAzY}. This process eliminated one of the $z\sim8$ galaxies.

Low-mass M, L and T dwarfs have spectra that peak in the near-infrared with a blue-ward drop off at $1~\mathrm{\mu m}$. Thus, they can mimic the spectra of Lyman-break galaxies at $z\sim8-9$. To minimise this type of contamination, in RB22, we fit all candidates with the \texttt{SpeX} prism library of brown dwarf spectral templates \citep{Burgasser2014}, discarding candidates if the resulting \texttt{EAzY} $\chi^2$ indicated a preferred fit compared to the best-fit galaxy spectrum. This step eliminated seven $z\sim8$ candidates and three $z\sim9$ candidates. 
After these steps, we have 32 candidates: 27 at $z\sim8$, three at $z\sim9$, and two at $z\sim10$.

Despite our best efforts to limit sample contamination, it is still possible that the final samples contain some degree of contamination by brown dwarfs. To add statistical robustness, we quantify the probability that the candidate is a galaxy (at any redshift) rather than a brown dwarf, $P_\textrm{gal}$, to be used to estimate the UVLFs in the following sections. We define $P_\textrm{gal}$ as a combination of a Bayesian Information Criterion (BIC) resulting from the \texttt{EAzY} fits and a scaling fraction denoting the relative surface area density of galaxies versus brown dwarfs in our observations: 
\begin{equation}
\begin{gathered}
    P_\textrm{gal} = a \times  \frac{\textrm{BIC}_\textrm{BD}}{\textrm{BIC}_\textrm{gal}}\times\frac{\textrm{Prior}_\textrm{gal}}{\textrm{Prior}_\textrm{BD}}\\
    P_\textrm{BD} = a \times\frac{\textrm{BIC}_\textrm{gal}}{\textrm{BIC}_\textrm{BD}}\times\frac{\textrm{Prior}_\textrm{BD}}{\textrm{Prior}_\textrm{gal}}
\end{gathered}
\label{eq:pgal}
\end{equation}
BICs compare the log-likelihood of the best-fit galaxy spectrum to the log-likelihood of the best-fit brown dwarf spectrum, taking the number of parameters in the fit into account. The prior $\textrm{Prior}_\textrm{BD}$ is the number density of brown dwarfs over the observed area, while $\textrm{Prior}_\textrm{gal}$ is the number density of LBGs. The latter can technically include low-z interlopers with which we deal later. The normalisation constant $a$ is derived by imposing the condition $P_\textrm{gal}+P_\textrm{BD}=1$. 

RB22 used the number density of brown dwarfs from \citet{Aganze2022}, namely 164 ultra-cool dwarfs in 0.6 deg$^2$ as the $\textrm{Prior}_\textrm{BD}$ and the number of candidates or 32 galaxies per total observing area for the $\textrm{Prior}_\textrm{gal}$. Using Equation \ref{eq:pgal} to calculate $P_\textrm{gal}$, they found that one $z\sim8$ candidate had $P_\textrm{gal}$ value equal to 0.4, indicating that it is likely a brown dwarf, and discarded it from the sample. Therefore, the final sample from RB22 consists of 31 galaxies: 26 at $z\sim8$, three at $z\sim9$, and two at $z\sim10$. The candidates, as well as their photometric information, are listed in Table A1 of RB22. These candidates constitute the sample for the UVLFs calculation in this paper. 

As an improvement over RB22, we reestimate the $P_\textrm{gal}$ values using more detailed priors. Both $\textrm{Prior}_\textrm{BD}$ and $\textrm{Prior}_\textrm{gal}$ are now magnitude dependent. Additionally, $\textrm{Prior}_\textrm{BD}$s are  field-specific. We start with the $\textrm{Prior}_\textrm{BD}$ term by following\citet{Aganze2022b} to calculate the number density of M7-T8 stars at the coordinates of our candidates. The model assumes that brown dwarfs reside in three locations: the thin disk, the thick disk and the halo of the Milky Way. The thin-disk vertical scaleheights are specific to the spectral type of the brown dwarfs, ranging from 146 pc to 249 pc. The thin disk radial scaleheight, the thick disk radial scaleheight and the thick disk vertical scaleheight are constants: 2600 pc, 3600 pc, and 900 pc respectively. The halo is assumed to have a flattened spheroid density population with parameters listed in \citet{Aganze2022b}. For each dwarf spectral type, we calculate the minimum and maximum distance ($d_\textrm{min}$, $d_\textrm{max}$) such that the dwarfs would have apparent magnitudes within $\pm0.5$ mag of the candidate's F160W magnitude. We then calculate the number of brown dwarfs with the spectral type SPT in 1 arcmin$^2$ at the location of the candidate's galactic position,
\begin{equation}
N(\textrm{SpT}) = C\times\Delta \Omega \int_{d_\textrm{min}(\textrm{SpT})}^{d_\textrm{max}(\textrm{SpT})} \rho(\mathbf{r}) r^2dr
\label{eq:Prior_BD}
\end{equation}
The density, $\rho(\mathbf{r})$, is the combination of the number densities from the thick disk, thin disk and halo, i.e. $\rho(\mathbf{r}) = \rho_\mathbf{\textrm{thin}}(\mathbf{r})+\rho_{\textrm{thick}}(\mathbf{r})+\rho_{\textrm{halo}}(\mathbf{r})$. They are functions of galactocentric position vector $\mathbf{r}$. The integration is along the geocentric distance r. $\Delta\Omega$ is the solid angle in steradian equal to 1 arcmin$^2$. We assume that the dwarfs have the same completeness $C$ as LBGs. Therefore $C$ is the completeness derived in Section \ref{sec:completeness}, e.g. the value in Figure \ref{fig:completeness_Minput_example} at the redshift and MUV bin of the candidate. Finally, the prior $\textrm{Prior}_\textrm{BD}$ is the sum of the $N(\textrm{SpT})$ for all spectral types from M7 to T8.

As for the $\textrm{Prior}_\textrm{gal}$ term, the number density of LBGs is rather uncertain as it is something we are trying to constrain. We only know the maximum number in the case of no brown dwarf contamination, which is the current number of candidates in the magnitude bin of interest (within $\pm0.5$ mag of the candidate's F160W magnitude) divided by the survey area in Table \ref{tab:field_summary}. However, we expect that some of these are contaminants. In literature, most studies only discard photometric-selected LBG candidates whose SEDs have worse fits with galaxy templates than with cool dwarf templates and do not further estimate $P_\textrm{gal}$ \citep[e.g.,][]{Stefanon2017b,Rojas-Ruiz2020}. However, those that attempted to quantify the contamination by dwarfs generally estimate the contamination by brown dwarfs to be $\sim20-25\%$. \citet{Calvi2016} estimated dwarf contamination rates in the early BoRG survey from the number of archival stars that may enter their LBG selection box. They estimated that one in four (25\%) of their $z\sim7.5$ candidates may be a cool dwarf instead. \citet{Kauffmann2022} perturbed the photometry of their ground-based selected candidates with noise and repeated the $\chi^2$ comparison between the galaxy and cool dwarf templates. They found that three out of fifteen $z\sim8$ candidates (20\%) may be potential cool dwarfs. 
Therefore, we incorporate these literature findings into the $\textrm{Prior}_\textrm{gal}$ term. We multiply the current number density of candidates in the magnitude bin of interest by 0.75 to obtain $\textrm{Prior}_\textrm{gal}$, i.e. we assume that 25\% are generally contaminants. 

We now proceed to calculate $P_\textrm{gal}$ with Equation \ref{eq:pgal} using the $\textrm{Prior}_\textrm{gal}$ and $\textrm{Prior}_\textrm{BD}$ calculated above. They are listed in Table \ref{table:sample}. These $P_\textrm{gal}$ values are magnitude dependent and field-specific. The $1\sigma$ range for $z\sim8$ candidates is from 0.64 to 0.99. Candidate 1917-3335\_929 has the least $P_\textrm{gal} = 0.11$ because its galactic coordinate is in the direction of the galactic centre ($l=+4.4\degree,b=-19.6\degree$). $z\sim9-10$ candidates all have very small chance of being dwarfs (<3\%) because the SEDs of the galaxies at these redshifts are not well fit by brown dwarf templates. We include these $P_\textrm{gal}$ probabilities in the calculation of UVLFs in Section~\ref{sec:UVLF}.


Even if the candidates are determined to be galaxies (i.e. not brown dwarfs), it is also possible that some are low-z interloping galaxies. These interlopers are likely evolved $z\sim2$ galaxies with a strong Balmer break that may be mistaken as a Lyman break of $z\gtrsim8$ galaxies. Here, we quantify the average number of possible interlopers specific to each observed field. The advantage of quantifying the average number of interlopers over quantifying the interloper fraction is that it only relies on the known properties of galaxies at low redshifts. It does not rely on the number density of $z\gtrsim8$ galaxies, which is something we are trying to constrain. To do so, we use the JAdes extraGalactic Ultradeep Artificial Relizations (JAGUAR) v1.2 catalogues of $0.2<z<15$ galaxies \citep{Williams2018}. The galaxies in the JAGUAR catalogues include both quiescent and star-forming galaxies that follow observed properties of galaxies from the current surveys, such as stellar mass functions, \MUV–$M_*$ relations and the $\beta$-\MUV\ relations. Their spectral model includes emission and absorption from stellar, photoionised gas and dust components. We used all ten realisations of the provided photometric catalogues, equivalent to an area of 1210 arcmin$^2$. From the catalogues, we consider the galaxies at $1<z<4$ as potential interlopers and take their intrinsic photometric information. Since the JAGUAR catalogues do not provide the photometry of all the bands observed in the SuperBoRG, we calculate the magnitudes for those missing bands (e.g. F600LP, F300LP, and F475W) from their spectral catalogues directly. We then add noise to the photometry according to the depths of the SuperBoRG fields. We then pass these noise-added fluxes through the candidate selection described in Section \ref{sec:dropout_selection} and Section \ref{sec:photometric_redshift}. For each field, we obtain the number density of the low-z galaxies that pass the LBG selection criteria and normalise it with the area associated with the field to obtain the expected number of interlopers in the field ($\overline{N}_\textrm{interlopers}$). They are also listed in Table \ref{table:sample}. For $z\sim8$ selection criteria, all but one field have $\overline{N}_\textrm{interlopers}$ smaller than 0.3 galaxies. The field 0728+0509 has $\overline{N}_\textrm{interlopers}$ of 2.3 galaxies. Such a high number is due to the shallow exposure in the non-detection F814W band, whose $5\sigma$ limiting magnitude is 24.8 mag, which is almost two magnitudes shallower than typical depths of non-detection bands in other fields. The median and standard deviation of $\overline{N}_\textrm{interlopers}$ excluding this outlier field are 0.13 and 0.09 galaxies, respectively. For $z\sim9$ and $z\sim10$ selection criteria, the expected number of interlopers are much smaller, at lower than 0.02 galaxies per field. This is likely because, at higher redshifts, we have more available non-detection bands that can effectively filter out potential interlopers. 

If we assume that the chance that a candidate is a true high-z galaxy equals to $P_\textrm{gal}\times(1-\overline{N}_\textrm{interlopers})$, i.e., not brown dwarfs and not low-z interlopers, our candidates have median probability of 0.81 and $1\sigma$ range of (0.58, 0.95) of being true high-z galaxies. Regardless, we use $P_\textrm{gal}$ and $\overline{N}_\textrm{interlopers}$ independently for the UVLF calculation presented in Section~\ref{sec:UVLF}. Even though individual galaxies have a certain probability of being a contaminant, we statistically account for it so that the UVLF determination is robust.

\begin{table}
\begin{center}
\renewcommand{\arraystretch}{1.3}
\begin{tabular}{l c c c c} 
 \hline
 Field\_ID & $z_\textrm{phot}$ & $M_\textrm{UV}$ & $P_\textrm{gal}$ & $\overline{N}_\textrm{interlopers}$ \\ 
 \hline\hline
0314-6712\_383 & 7.75 & -23.10 & 0.52 & 0.00 \\
0409-5317\_313 & 8.02 & -20.90 & 0.92 & 0.00 \\
0440-5244\_742 & 7.84 & -21.37 & 0.96 & 0.00 \\
0830+6555\_244 & 7.75 & -21.60 & 0.90 & 0.13 \\
0853+0310\_112 & 7.66 & -22.18 & 0.99 & 0.03 \\
0925+1360\_899 & 7.58 & -21.38 & 0.99 & 0.18 \\
0948+5757\_697 & 7.66 & -21.33 & 0.98 & 0.17 \\
0955+4528\_914 & 8.20 & -21.18 & 0.99 & 0.20 \\
0956+2848\_986 & 8.11 & -20.53 & 0.99 & 0.06 \\
1017-2052\_310 & 8.11 & -21.10 & 0.99 & 0.31 \\
1033+5051\_164 & 7.93 & -21.28 & 0.99 & 0.00 \\
1104+2813\_447 & 7.58 & -21.37 & 0.99 & 0.13 \\
1218+3008\_638 & 7.58 & -21.17 & 0.97 & 0.02 \\
1437+5043\_1241 & 7.84 & -21.41 & 0.99 & 0.01 \\
1515-1517\_698 & 7.66 & -21.20 & 0.58 & 0.18 \\
1558+0812\_601 & 7.66 & -22.05 & 0.89 & 0.17 \\
1917-3335\_929 & 8.11 & -21.06 & 0.11 & 0.14 \\
2203+1851\_1071 & 7.84 & -21.58 & 0.88 & 0.00 \\
0728+0509\_232 & 8.20 & -22.14 & 0.32 & 2.27 \\
0104+0021\_339 & 7.66 & -21.75 & 0.98 & 0.02 \\
1149+2202\_169 & 7.66 & -21.47 & 0.98 & 0.27 \\
1149+2202\_343 & 8.38 & -22.13 & 0.87 & 0.27 \\
2134-0708\_2928 & 7.58 & -21.18 & 0.72 & 0.17 \\
0859+4114\_138 & 7.75 & -21.84 & 0.64 & 0.20 \\
0859+4114\_718 & 7.75 & -21.93 & 0.87 & 0.20 \\
1115+2548\_455 & 8.38 & -21.26 & 0.98 & 0.05 \\
\hline
1607+1332\_996 & 8.76 & -21.87 & 0.97 & 0.00 \\
0037-3337\_563 & 8.76 & -21.41 & 0.99 & 0.02 \\
1420+3743\_1025 & 8.76 & -21.46 & 0.99 & 0.00 \\
\hline
1459+7146\_344 & 10.00 & -21.59 & 0.99 & 0.00 \\
1142+3020\_67 & 9.79 & -22.36 & 0.99 & 0.00 \\
\hline
\end{tabular}
\end{center}
\caption{The sample in this work, taken from RB22. $P_\textrm{gal}$ is the probability that the galaxy is not a brown dwarf. The two numbers refer to the low and high contamination-by-dwarfs scenario, respectively. $\overline{N}_\textrm{interlopers}$ is the estimated number of interlopers in the field (see Section \ref{sec:potential_contamination}). We refer the reader to Table 7 in RB22 for other details on the candidates, such as the coordinates, photometry, etc.}
\label{table:sample}
\end{table}

\subsubsection{Other sources of uncertainty that are not accounted for}
To ensure our determinations of the UVLF are not significantly affected by potential lensing effects that weren't previously accounted for, here we follow the procedure of \citet{Mason2015} to identify whether any of our candidates experience multiple imaging or an intermediate magnification. We first use \texttt{EAzY} to determine the redshifts of objects in the vicinity of each candidate, using the available HST bands and a flat prior. If there is any $z<3$ source that is less than 18 arcsec away, we consider it a potential deflector. We then estimate its velocity dispersion by using the velocity dispersion -- apparent magnitude relation (as a function of redshift) provided by \citet{Mason2015}. The magnification that the dropout experiences due to this deflector is equal to 
\begin{equation}
    \mu = \frac{|\theta|}{|\theta|-\theta_\textrm{ER}},
\end{equation}
where $\theta$ is the angular separation between the lens and the source in the image plane. $\theta_\textrm{ER}$ is the Einstein radius of the lens. Assuming that the lens is a singular isothermal sphere, 
\begin{equation}
     \theta_\textrm{ER}=4\pi\frac{D_{ls}}{D_s}\Big\{\frac{\sigma}{c}\Big\}^2.
\end{equation}
$D_{ls}$ and $D_s$ are the angular diameter distances between the lens and the source and between the observer and the source, respectively. $\sigma$ is the velocity dispersion and $c$ is the speed of light. 

Based on the procedure above, we consider that the dropout is lensed if there is a foreground object that produces a mean magnification larger than $\mu=1.4$. Most of the foreground objects are sufficiently far from our candidates. They produce mean magnification $\mu\approx1.0$, i.e. no magnification. Candidate 0956+2848\_986 is the only candidate that may experience some gravitational lensing with a magnification of $\mu=1.5\pm1.3$ by a $z\sim1.3$ foreground galaxy located $\sim 3$ arcsec away. This candidate happens to be the faintest in our sample. Its UV magnitude before magnification correction is $-20.5$ mag, and correcting the flux of this galaxy with the derived magnification does not significantly change the derived UVLFs. Thus, we do not correct the UVLFs for the effect of gravitational lensing in the following sections.

\subsection{UV magnitude measurements}

\begin{figure}
    \centering
    \includegraphics[width=\textwidth]{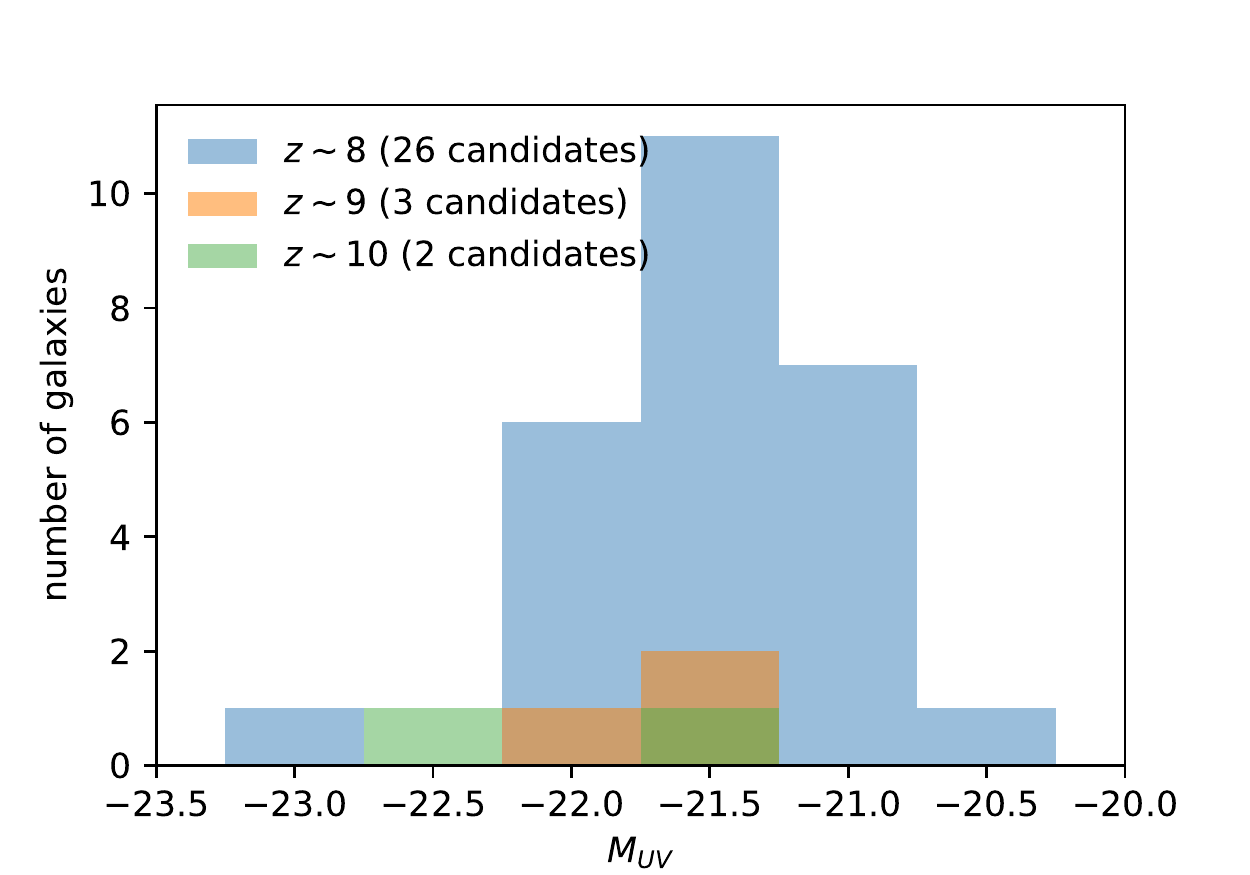}
    \caption{Histogram of UV magnitudes for $z=8-10$ candidates in this study.}
    \label{fig:number_candidates}
\end{figure}

In this work, we assume the values of intrinsic UV magnitudes of the candidates, as we reported in RB22. They were derived with the SED-fitting code \texttt{Bagpipes} \citep{Carnall2018}, with redshifts fixed to values provided by \texttt{EaZY}. The SFH is assumed to be exponentially declining. The intrinsic UV magnitudes are measured at $1600$ \angstrom, defined as an average of the spectra in the rest-frame $1550-1650$ \angstrom. Figure \ref{fig:number_candidates} shows the histogram of the measured UV magnitudes of the candidates. The SED fitting code \texttt{Bagpipes} was used instead of \texttt{EaZY} because it assumes more physically-motivated SFHs in the fitting procedure. We have crossed check with the UV magnitudes measured with \texttt{EaZY}. All \MUV\ values are consistent within 0.1 mag.

\section{Completeness simulations}
\label{sec:completeness}

We use \texttt{GLACiAR2} \citep{Carrasco2018, Leethochawalit2022} to perform the injection-recovery simulation and calculate the completeness function for every SuperBoRG field. For each field, we inject galaxies in bins of intrinsic redshift and \MUV. The redshift bins have increments of $\Delta z = 0.25$ and range from $z=7$ to $z=9$ for the $z\sim8$ dropout selection (9 redshift bins), and from $z=8$ to $z=12.5$ for the $z\sim9$ and $z\sim10$ dropout selections (19 redshift bins). The \MUV\ bins range from $-23.5$ mag to $-20.0$ mag with an increment of 0.5 mag (8 magnitude bins). For each bin of redshift and \MUV, we randomly draw 40 \MUV\ values according to the probability described by the Schechter functions whose parameters derived by \citet{Schmidt2014} for the $z\sim8$ galaxies and by \citet{Morishita2018} for $z>8$ galaxies. The pairs of \MUV\ and redshift values constitute 2880 iterations for the $z\sim8$ dropout simulation and 6080 for the $z\sim9$ and $z\sim10$ dropout simulation.  

We generate a library of high-redshift star-forming galaxy spectra to be used as the injected galaxies' spectra. To create the library, we randomly select 1000 spectra of $z>6$ star-forming galaxies from the JAGUAR \citep[JAGUAR][]{Williams2018} catalogues. These spectra were generated with the \texttt{BEAGLE} tool \citep{Chevallard2016}. The galaxies follow the observed fundamental metallicity relation and the metallicity-ionization parameter ration, have delayed SFHs, and are accounted for dust attenuation. We set this spectral library as an input for \texttt{GLACiAR2}. For each \texttt{GLACiAR2} iteration (one \MUV\ value and one redshift value), we inject 200 galaxy stamps into each SuperBoRG field. For each iteration, we derive the photometry of all galaxies from a randomly drawn spectrum using the spectral library described above. The spectrum is normalised to match the required \MUV, and red-shifted to match the input redshift $z$. We then calculate the apparent magnitudes of the mock galaxies given the response functions of the HST filters. We assign a S\'ersic brightness profile ($n=1$) with a random inclination and ellipticity to each simulated galaxy. The choice for the S\'ersic index $n=1$ is suitable for the typical clumpy morphology of high-z galaxies \citep[e.g.,][]{Shibuya2016}. We set the half-light radius of each galaxy to inversely scale with redshift $(1+z)^{-1}$, where we fix the half-light radius of a $z=8$ galaxy to 0.7 kpc, equal to the median half-light radius of our highest-confidence $z\sim8$ SuperBorG candidates \citep{Roberts-Borsani2021}. 

\texttt{GLACiAR2} uses \texttt{SExtractor}\citep{BertinArnouts1996} to recover the injected galaxies. Therefore, in the completeness simulations, we use the same parameter files and set detection images used for candidate selections in Section \ref{sec:data_and_sample_selection} (i.e. F125W detection for $z\sim8$ dropouts and F140W+F160W for $z>8$ dropouts). 

We consider the injected galaxies as recovered when classified as `isolated' or `minimally blended' by \texttt{GLACiAR2}, i.e. those with detection status $\geq 0$. We then pass the recovered photometries to the dropout selection and \texttt{EAZY} photometric-redshift code.  The simulated sources are deemed to be dropout candidates if they satisfy the photo-z probability distribution requirements used for the actual source selection. This way, our source recovery selection replicates exactly all steps used for source selection, excluding the visual selection for artefact identification. Similar to the procedure in Section \ref{sec:data_and_sample_selection}, the recovered \MUV\ magnitudes of the recovered galaxies are the average fluxes in the rest-frame $1550-1650$ \angstrom\  wavelength range of their best-fit SEDs.

\subsection{Completeness functions and effective volumes}
\label{sec:Completeness_functions_and_effective volumes}
\begin{figure}
    \centering
    \includegraphics[width=0.95\textwidth]{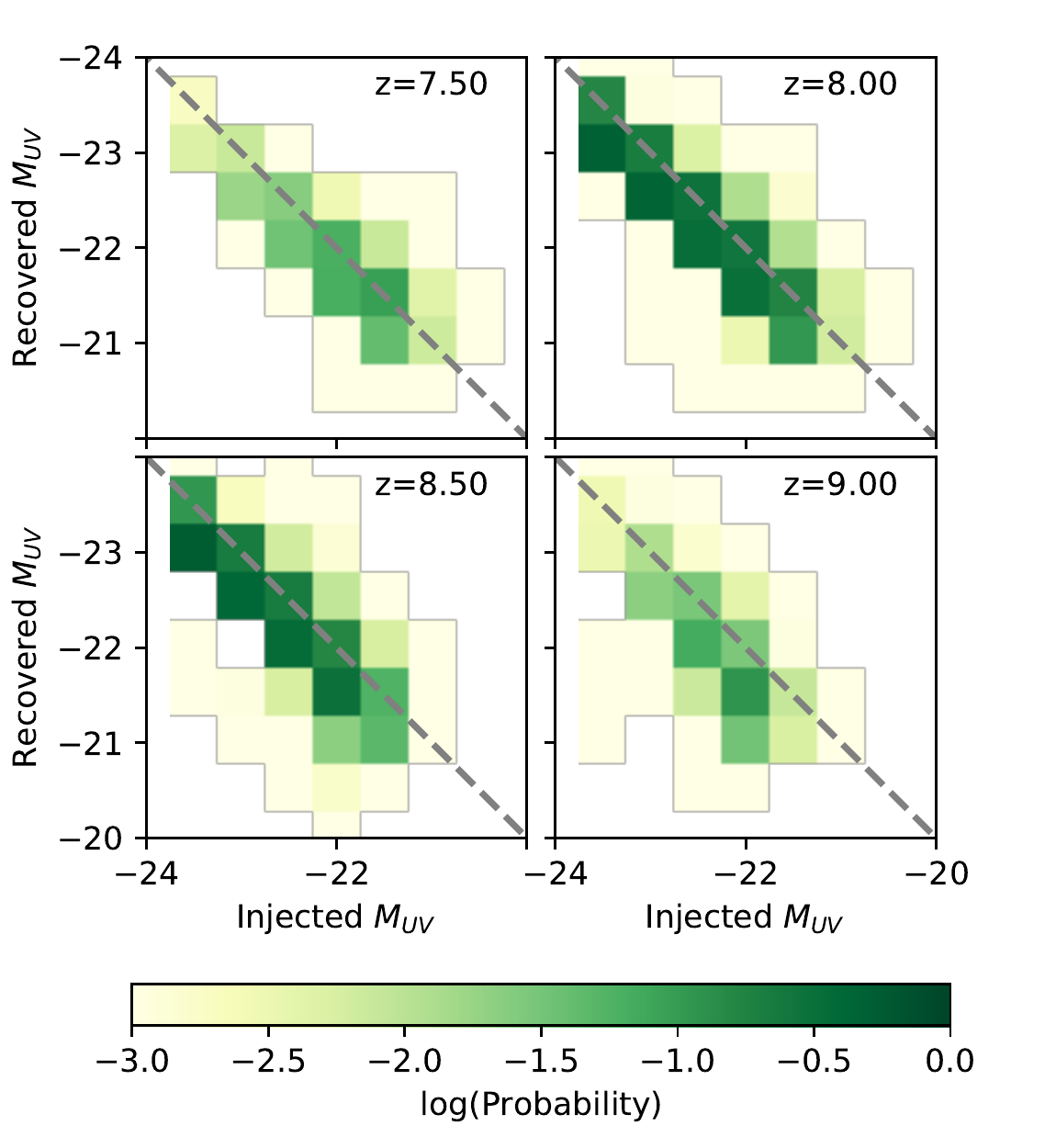}\\
    \caption{Example of a completeness function as a function of intrinsic \MUV\ and recovered \MUV\ for the $z\sim8$ selection criteria. We show the results for the BoRG 0834+5238 field, which is a typical field in the SuperBoRG project. The injected redshift of the galaxies is marked at the top right corner of each panel. }\label{fig:completeness_MinputMoutput_example}
\end{figure}
\begin{figure*}
    \centering
    \includegraphics[width=0.31\textwidth]{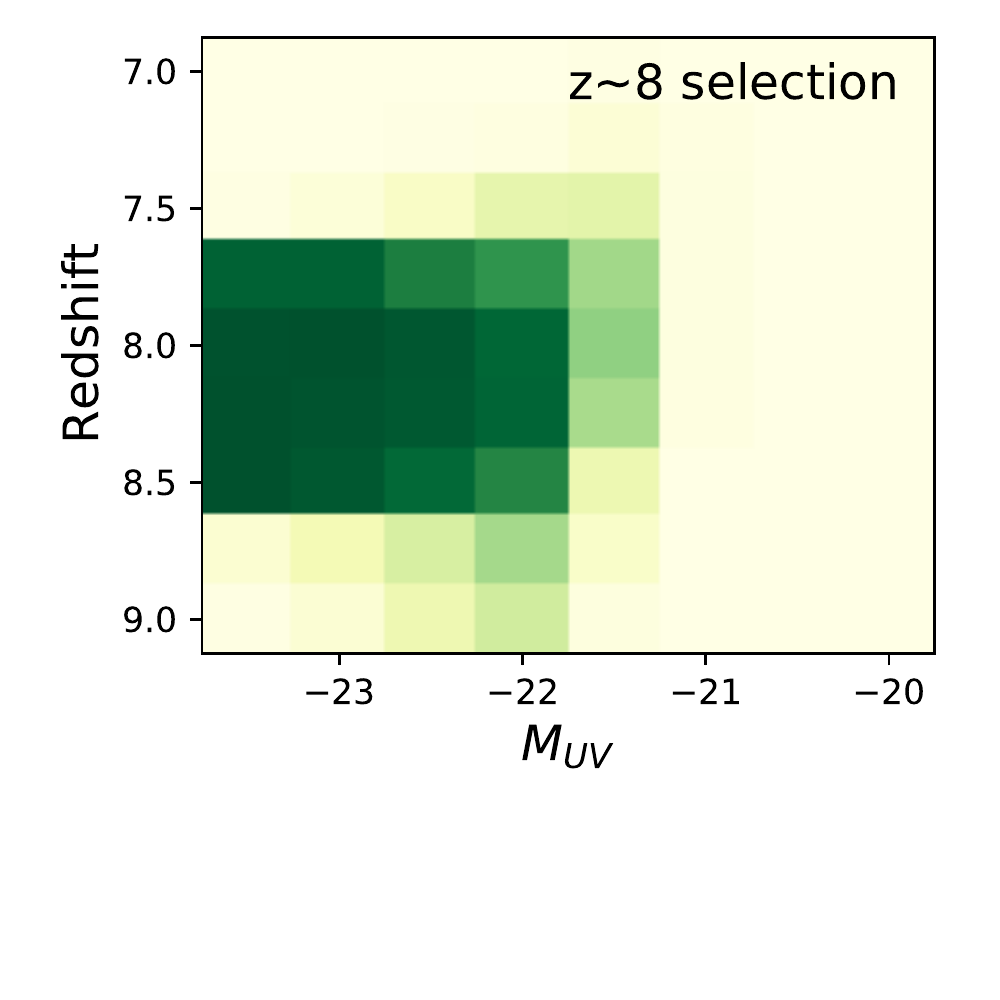}
    \includegraphics[width=0.31\textwidth]{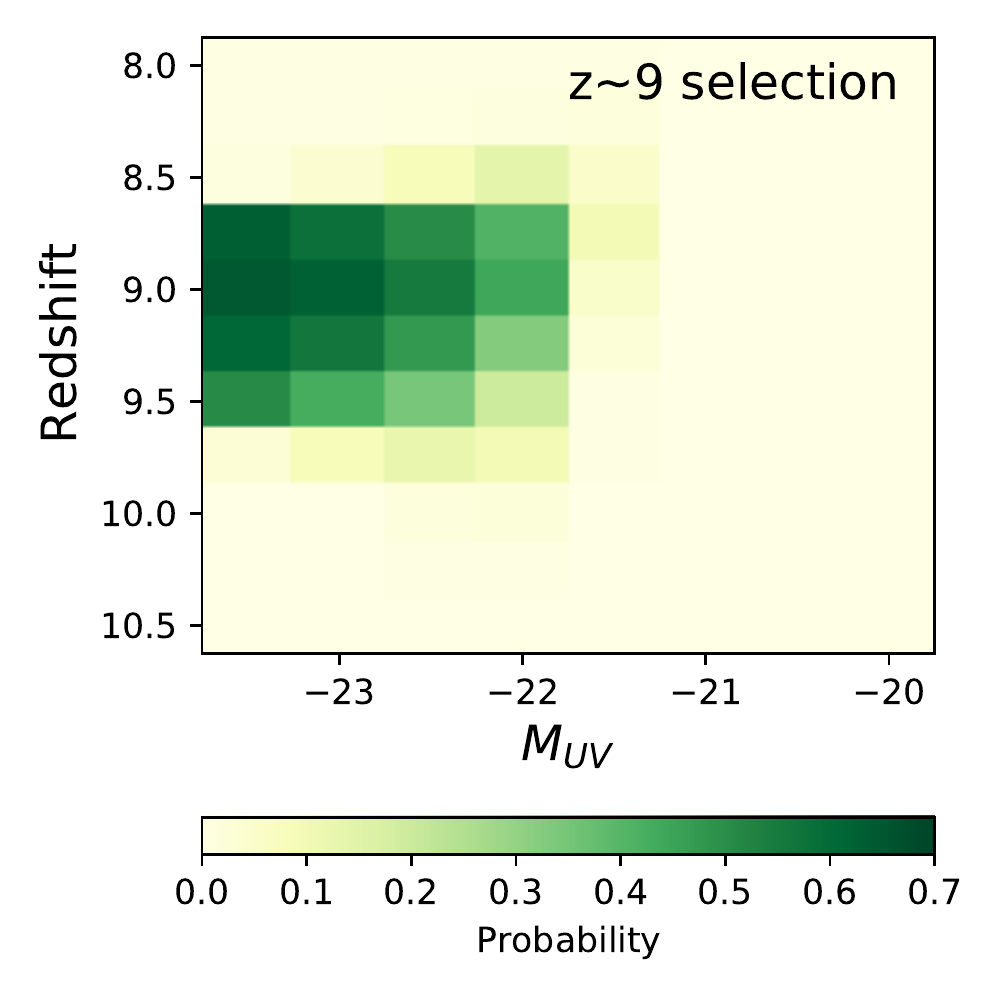}
    \includegraphics[width=0.31\textwidth]{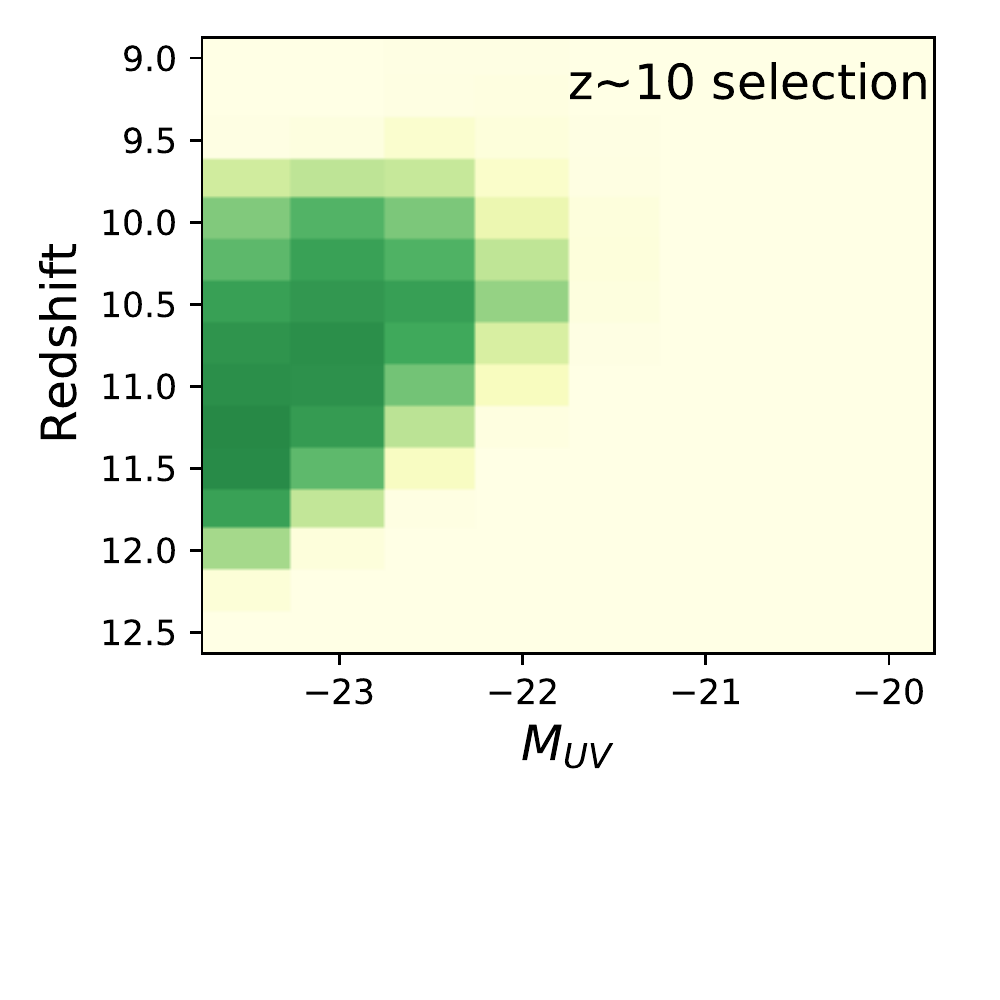}
    \caption{Example of completeness functions as a function of intrinsic magnitude and redshift. As in previous figure, we show the results for BoRG 0834+5238 field.}
    \label{fig:completeness_Minput_example}
\end{figure*}
Previous luminosity function studies construct completeness functions differently, despite using a similar concept of what a completeness simulation is. Definitions include completeness functions as a function of intrinsic \MUV\ \citep[e.g.][]{Finkelstein2016}, completeness as a function of recovered magnitudes \citep[e.g.][]{Oesch2007, Bowler2020}, and completeness as a function of both intrinsic and recovered \MUV\ \citep[e.g.][]{Bouwens2016}. As demonstrated in \citet{Leethochawalit2022}, the UVLF derived with a completeness function that is a function of both intrinsic and recovered UV magnitude is the least sensitive to flux scatter. Therefore, we will primarily employ this UVLF derivation method in this work, using it as our fiducial method. The completeness is thus defined as 
\begin{equation}
     P(M_\textrm{recov},M_\textrm{in},z)  = \frac{N(M_\textrm{recov},M_\textrm{in},z)}{N(M_\textrm{in},z)}
     \label{eq:completeness_function_min_mrecov_defn}
\end{equation}

$N(M_\textrm{in},z)$ is the number of injected galaxies in an intrinsic $M_\textrm{in}$ and $z$ bin. $N(M_\textrm{recov},M_\textrm{in},z)$ is the number of injected galaxies in the intrinsic $M_\textrm{in}$ and z bin but with $M_\textrm{recov}$ magnitudes. We show an example of such completeness function for $z=8$ galaxies of a SuperBorg field (BoRG0834+5238) in Figure \ref{fig:completeness_MinputMoutput_example}. The one-to-one relations are shown as dashed lines. It is clear that (1) there is flux scatter and (2) on average, the recovered \MUV\ are fainter than the intrinsic \MUV\ fluxes. We defer a detailed discussion on flux scatter to Section~\ref{sec:scatter} below. 

Although we derive the luminosity functions using the completeness defined above as our fiducial method, we also explore other commonly used definitions of completeness functions and their corresponding UVLF determinations (see Appendix \ref{appendix:other_methods}). A more intuitive type of completeness function, when plotted, is the completeness function as a function of intrinsic \MUV\ \citep[e.g.][]{Finkelstein2015}. It is the fraction of recovered galaxies at each injected \MUV and redshift bin. We show examples of this completeness function for the field BoRG0834+5238 in Figure \ref{fig:completeness_Minput_example}. The figure shows that our dropout selection criteria are reasonably effective at picking out the galaxies at the intended redshift ranges, despite the limited signal-to-noise ratios of the observations. Note that at the brightest magnitude bins, the completeness functions plateau at $\sim60\%$ for the $z\sim8-9$  selections and plateau at $\sim50\%$ for the $z\sim10$ selection. This is because of the successive steps (colour selection and photometric redshift criteria) we used to prevent significant contamination in the sample selection. In addition, foreground sources and artefacts in the non-dithered images affect a substantial amount of the image area. The variation (standard deviation) of the completeness among all the fields is $\sim10\%$, suggesting that the quality of the images across the sample is uniform.

\begin{figure}
    \centering
    \includegraphics[width=0.85\textwidth]{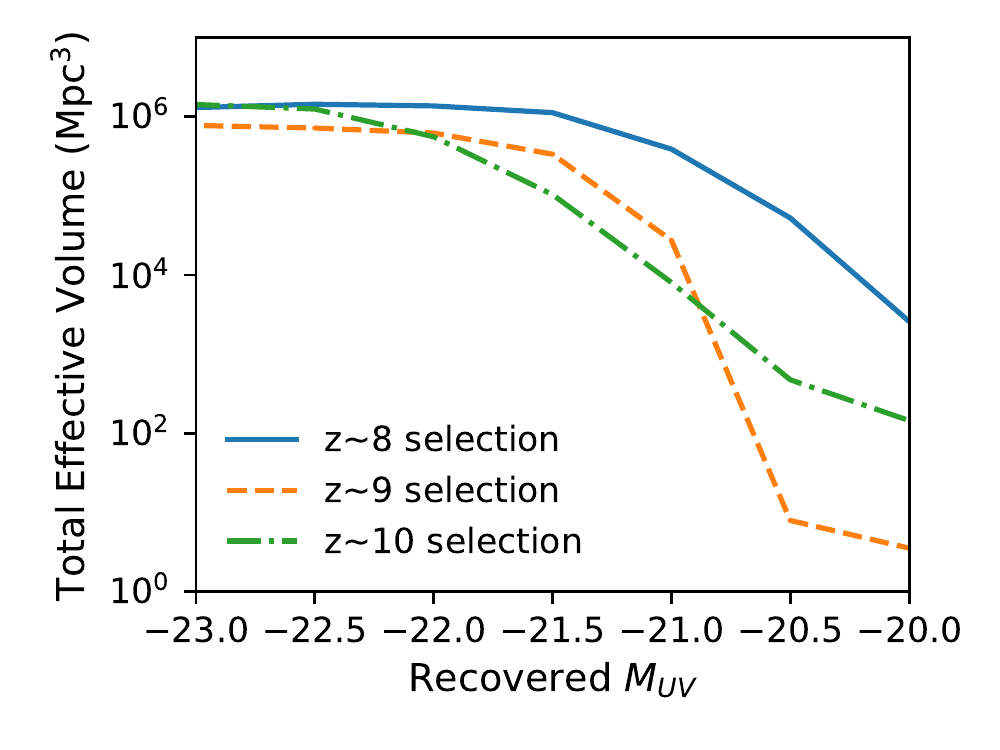}\\
    \caption{Effective volumes of the SuperBoRG, for selection criteria at different redshifts. }\label{fig:effective_volume}
\end{figure}

We also calculate the effective volume of the survey according to the methods used, e.g. in  \citet{Oesch2012, Morishita2018}. The effective volume is generally shown as a function of recovered UV magnitude. It is defined as
\begin{equation}
    V_\textrm{eff}(M_\textrm{recov}) = \int\frac{dV}{dz}P(M_\textrm{recov},z) dz.
    \label{eq:veff_mrecov}
\end{equation}
$P(M_\textrm{recov},z)$ is the number of simulated galaxies at redshift $z$ that are recovered to have UV magnitude in the bin $M_\textrm{recov}$ (regardless of their intrinsic UV magnitude), divided by the number of injected galaxies at redshift $z$ with intrinsic UV magnitude equal to $M_\textrm{recov}$. We plot the effective volumes for each redshift selection in Figure \ref{fig:effective_volume}. At the brightest magnitude bins, the effective volumes at all redshifts are approximately $10^6$ Mpc$^3$.

\subsection{Flux scatter based on the completeness simulation}\label{sec:scatter}

\begin{figure}
    \centering
    \includegraphics[width=0.85\textwidth]{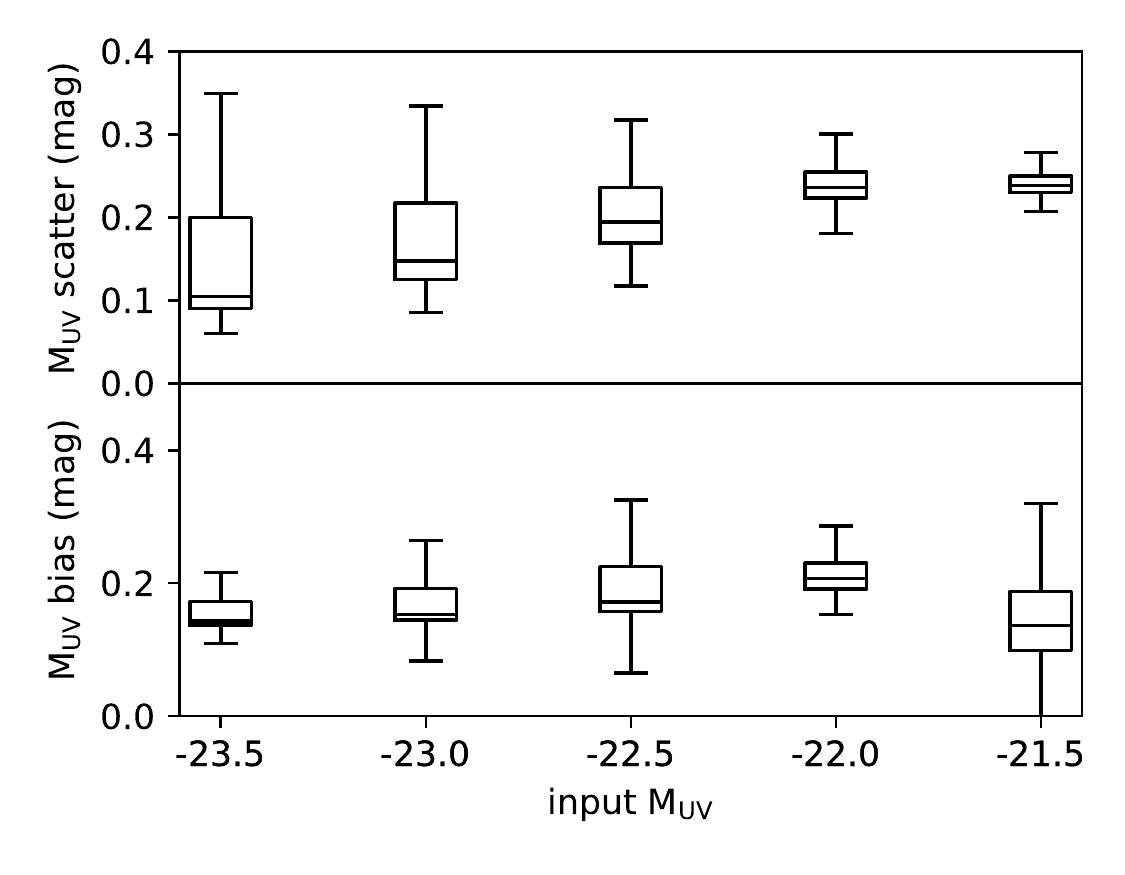}
    \caption{Flux scatter (upper panel) and flux bias (lower panel) of recovered galaxies in completeness simulations for the $z\sim8$ selection criteria. Each box plot shows the median, $1\sigma$ and $2\sigma$ lower and upper limits across all fields.  }\label{fig:fluxscatter}
\end{figure}

When the flux scatter is greater than 0.2 mag and if the flux scatter is not treated properly, the derived UVLFs can have a substantial bias at the bright end \citep{Leethochawalit2022}. Thus, we estimate the amount of flux scatter in our data from the completeness simulations to check whether flux scatter is of concern. We estimate the flux scatter of each field by comparing the injected UV magnitudes to the recovered UV magnitudes. We plot the results for the $z\sim8$ selection in Figure \ref{fig:fluxscatter}. The median flux scatters, shown in the upper panel, increase from $\sim0.1$ mag at the brightest magnitude bins to $\sim0.2$ mag around the magnitude bins at the drop off of the completeness. The distribution of the flux scatters across all fields generally skews right. We found that $\sim15\%$ of the fields have flux scatter larger than 0.2 mag, even at the brightest magnitude bins. These fields tend to be shallower fields observed by design with fewer NIR filters. 

As shown in the lower panel of Figure \ref{fig:fluxscatter}, the recovered absolute fluxes tend to be fainter than the intrinsic fluxes by $\sim0.2$ mag. This suggests that around $15-20\%$ of flux may be lost. Upon investigation, the most contributing factor is that SExtractor estimates Kron fluxes to be smaller than the input total fluxes.\footnote{The recovered redshifts estimated by \texttt{EAZY} normally agree with the input redshifts within a few percents (except at the injected bins of $z=9.3-9.5$ where the recovered redshifts are underestimated by $5-10\%$). Therefore, redshift uncertainty/bias does not overall contribute to the average absolute flux loss.} The amount of flux loss is consistent with the investigation by \citet{Merlin2019}, where they found that SExtractor could underestimate the total flux by $\sim15-20\%$ on average when a common and default value of the minimum Kron aperture radius, i.e. $R_\textrm{Kron,min}=3.5$, is used. To recover more than $95\%$ of the flux, they recommended $R_\textrm{Kron,min}$ to be $7-8$, which, in our opinion, is not commonly adopted as it risks the chance of having flux contaminated by other nearby sources. Another possible factor contributing to the underestimation is the strict Source Extractor parameter \texttt{DETECT\_THRESH} that we adopt to minimise the contamination from spurious hot pixels. As we directly adopt the sample from \citet{Roberts-Borsani2021} that was based on the photometric catalogue published by \citet{Morishita2021}, we use the same SExtractor parameters and selection methods in the completeness simulation for consistency. However, in this work, we implicitly account for the flux loss through the completeness correction when we fit for the best-fit UVLFs with the fiducial method. Therefore, any flux loss is not a concern for the UVLF determination.

\section{The UV luminosity function}
\label{sec:UVLF}
We derive parameterized UV luminosity functions using the method in \citet{Bouwens2015}, i.e. the fiducial method for completeness corrections as established in \citet{Leethochawalit2022}. In Appendix \ref{appendix:other_methods}, we also derive UVLFs using other methods commonly used in literature. We find that all methods yield consistent results, reassuring us of the robustness of the correction applied. The fiducial UVLF derivation is as follows. First, we calculate the total effective volume for the survey, which is a function of intrinsic magnitude $M_\textrm{in}$ and recovered magnitude $M_\textrm{recov}$. It is defined as
\begin{align} 
V_{\textrm{eff}}(M_\textrm{recov},M_\textrm{in}) &=  \sum_l V_{\textrm{eff},l}(M_\textrm{recov},M_\textrm{in}) \nonumber \\   &=  \sum_l \int P_l(M_\textrm{recov},M_\textrm{in},z) \frac{dV_C}{dz}\ \Omega_l\ dz,
\end{align}
where the summation runs over each field $l$ in the dataset (i.e. survey effective volume is the sum of the volumes of the individual fields). $P_l(M_\textrm{recov},M_\textrm{in},z)$ is the completeness function for field $l$ defined in Equation \ref{eq:completeness_function_min_mrecov_defn}. $V_C$ is the comoving volume per steradian at redshift $z$. $\Omega_l$ is the solid angle of the observed field $l$. With the effective volume, we can calculate the expected number of observed galaxies for a small bin of size $dM$ centred at $M_\textrm{recov}$ as a dot product between an effective comoving volume and the model UV luminosity function $\phi(M_\textrm{in})$:
\begin{equation}
     N_\textrm{model}(M_\textrm{recov}) =  V_\textrm{eff}(M_\textrm{recov},M_\textrm{in}) \cdot \phi(M_\textrm{in})dM.
     \label{eq:nexp_method3}
\end{equation}
We parameterise the UV luminosity function $\phi$ with a Schechter function 
\begin{align}
\phi(M_\textrm{in}|\alpha,M^*,\phi^*) =&\ 0.4 (\ln10) \phi^*\times10^{0.4(\alpha+1)(M^*-M\textrm{in})}\nonumber\\
& \times \exp{[-10^{0.4(M^*-M_\textrm{in})}]}
\end{align}
and with a double power-law function
\begin{align}
\phi(M_\textrm{in}|\alpha,\beta,M^*,\phi^*) = & 0.4 (\ln10) \phi^*\times[10^{0.4(\alpha+1)(M\textrm{in}-M^*)}\nonumber\\
& +10^{0.4(\beta+1)(M\textrm{in}-M^*)}].
\end{align}
Finally, we find the best-fit model parameters using a Markov chain Monte Carlo (MCMC) method via the python package \texttt{emcee} \citep{emcee}. We assume a Poissonian log-likelihood distribution:
\begin{equation}
     \ln(\mathcal{L}) = \sum_{M_\textrm{recov}}[N_{\textrm{obs}}(\ln N_\textrm{model})-N_\textrm{model}-\ln( N_\textrm{obs}!)].
\label{eq:poisson}
\end{equation}
Since our data is limited to the bright end of the UVLFs, we fix the faint-end slope $\alpha$ parameter (both Schechter and DPL fit) to the values derived by \citet{Bouwens2021}. $\alpha = -2.23, -2.33$, and -2.38 for the functions at $z\sim 8, 9$, and 10 respectively. These faint-end slopes are measured from deep HST imaging (e.g. HUDF, HFF, and HDUV fields), which reach \MUV$\sim-18$ mag at the faint end. 

As discussed in Section \ref{sec:potential_contamination}, each of our final candidates has a small quantifiable probability of being a contaminant. To compensate for this, we use the Monte Carlo method to iteratively resample the candidates. In each Monte Carlo iteration, we draw a subset of the sample according to their probabilities of being true high-z galaxies and run the MCMC described in the paragraph above. The drawing was done in two steps. First is according to the candidate's $P_\textrm{gal}$ calculated in Section \ref{sec:potential_contamination} i.e. the probability that they are not brown dwarfs. The second drawing is according to the candidates' probability of being a true high-z galaxy using the ${N}_\textrm{interlopers}$ derived earlier. If there is only one candidate in the HST field, then the chance that the candidate is a high-z galaxy is equal to the chance that there is no low-z interloper in the field. In this case, if we assume a Poisson distribution $\mathrm {Pois}(k;{N}_\textrm{interlopers})$ for the interlopers, then the chance is equal to $\mathrm{Pois}(k{=}0) = e^{-\overline{N}_\textrm{interlopers}}$. If there are two candidates in a field, which apply to two fields and four candidates in our sample, we perform the following. If both galaxies in the field are not brown dwarfs in the iteration, we determine case-by-case whether the two galaxies are low-z interlopers. We generate a random uniform number which ranges from 0 to 1. If the number is smaller than or equal to the chance of having no interlopers in the field, i.e. $\leq \mathrm{Pois}(k{=}0)$, then both candidates are considered to be true high-z galaxies. If the number is greater than $\mathrm{Pois}(k{=}0)$ but smaller than $\mathrm{Pois}(k{\leq}1) = e^{-\overline{N}_\textrm{interlopers}}+\overline{N}_\textrm{interlopers}e^{-\overline{N}_\textrm{interlopers}}$, then one of the two galaxies is randomly selected to be an interloper. Otherwise, both galaxies are considered to be interlopers and are not included in the sample of that iteration. We repeat the sample drawing and the fitting procedures using 100 Monte Carlo iterations.

\begin{figure}
    \centering
    \includegraphics[width=\textwidth]{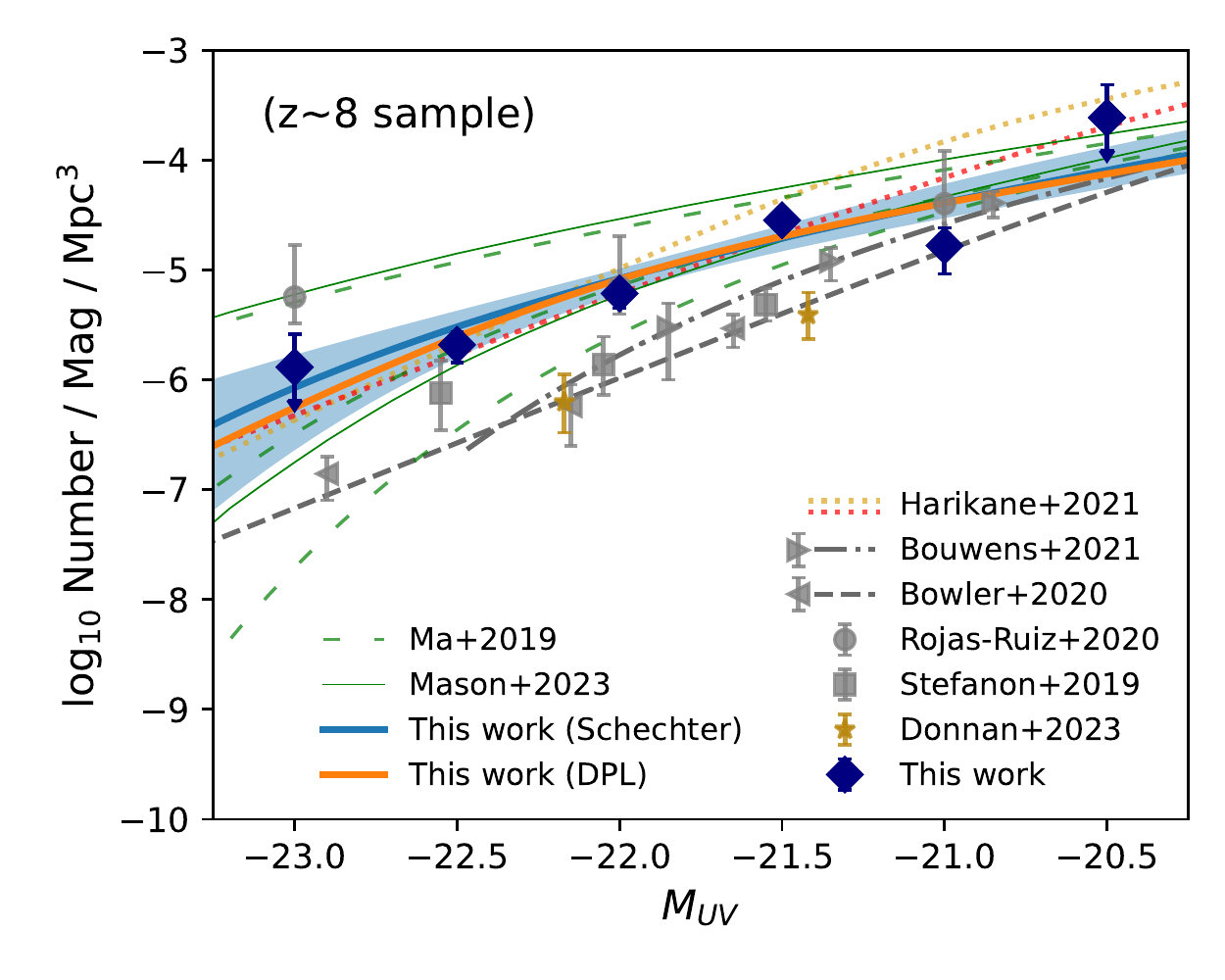}
    \caption{Derived UVLFs at $z=8$. The best-fit Schechter and DPL functions are shown in solid blue and orange lines. We only show the $1\sigma$ uncertainty range for the Schechter function in the blue shade. The size of $1\sigma$ uncertainty for the best-fit DPL function is not shown here but is comparable. The navy diamonds show nominal number densities (see Appendix \ref{appendix:other_methods}). Other observational constraints from ground-based and HST observations are shown in grey: circles - \citet{Rojas-Ruiz2020}, squares - \citet{Stefanon2019}, left triangles with a dashed-dotted line - \citet[DPL function]{Bowler2020}, and right triangles with dashed line \citet[Schechter function]{Bouwens2021}. Gold stars show early results of JWST observations \citep{Donnan2023}. The yellow and red dotted lines represent respectively the UVLFs at $z=6$ and $z=7$ of Lyman dropouts, which include both galaxies and AGNs, derived from Subaru/Hyper Suprime-Cam survey \citep{Harikane2021}. They show no evolution in the bright end at $z=6-7$. The green lines show theoretical predictions. From top to bottom, green loosely-dashed lines show predicted Schechter functions based on the FIRE-2 simulation with an assumed dust-to-metal ratio of 0.0, 0.4, and 0.8, respectively \citep{Ma2019}. The upper and lower green solid lines show the predictions from semi-empirical models with and without dust attenuation, respectively, by \citep{Mason2023}.}
    \label{fig:derived_UVLFs_z8}
\end{figure}

\begin{figure}
    \centering
    \includegraphics[width=\textwidth]{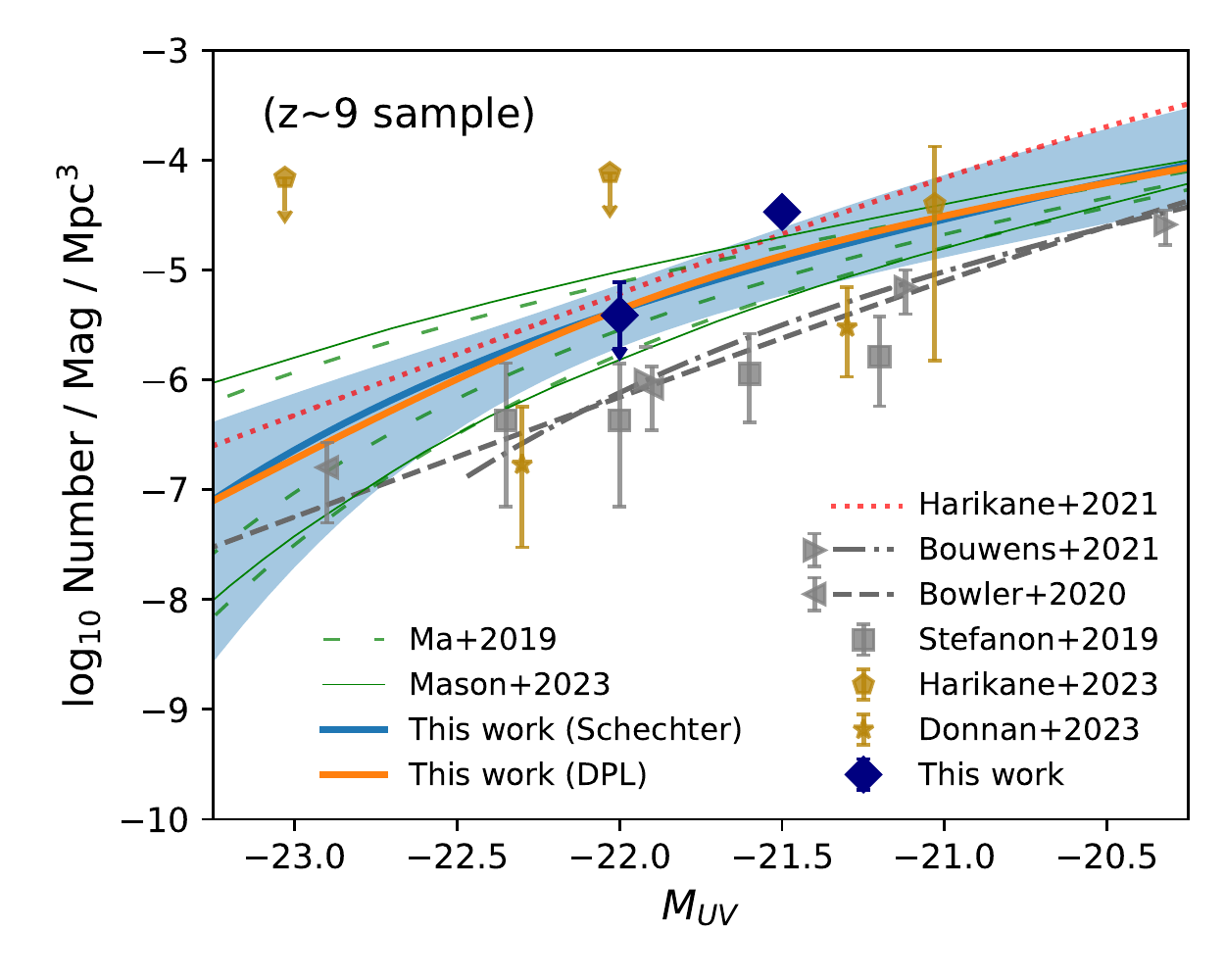}
    \caption{Similar to Figure \ref{fig:derived_UVLFs_z8} but for UVLFs at $z=9$. Here we only show the $z=7$ UVLF from \citet{Harikane2021} for reference. Gold pentagons show UVLFs determined from the latest data of JWST's ERO and ERS programs \citet{Harikane2023}.}
    \label{fig:derived_UVLFs_z9}
\end{figure}

\begin{figure}
    \centering
    \includegraphics[width=\textwidth]{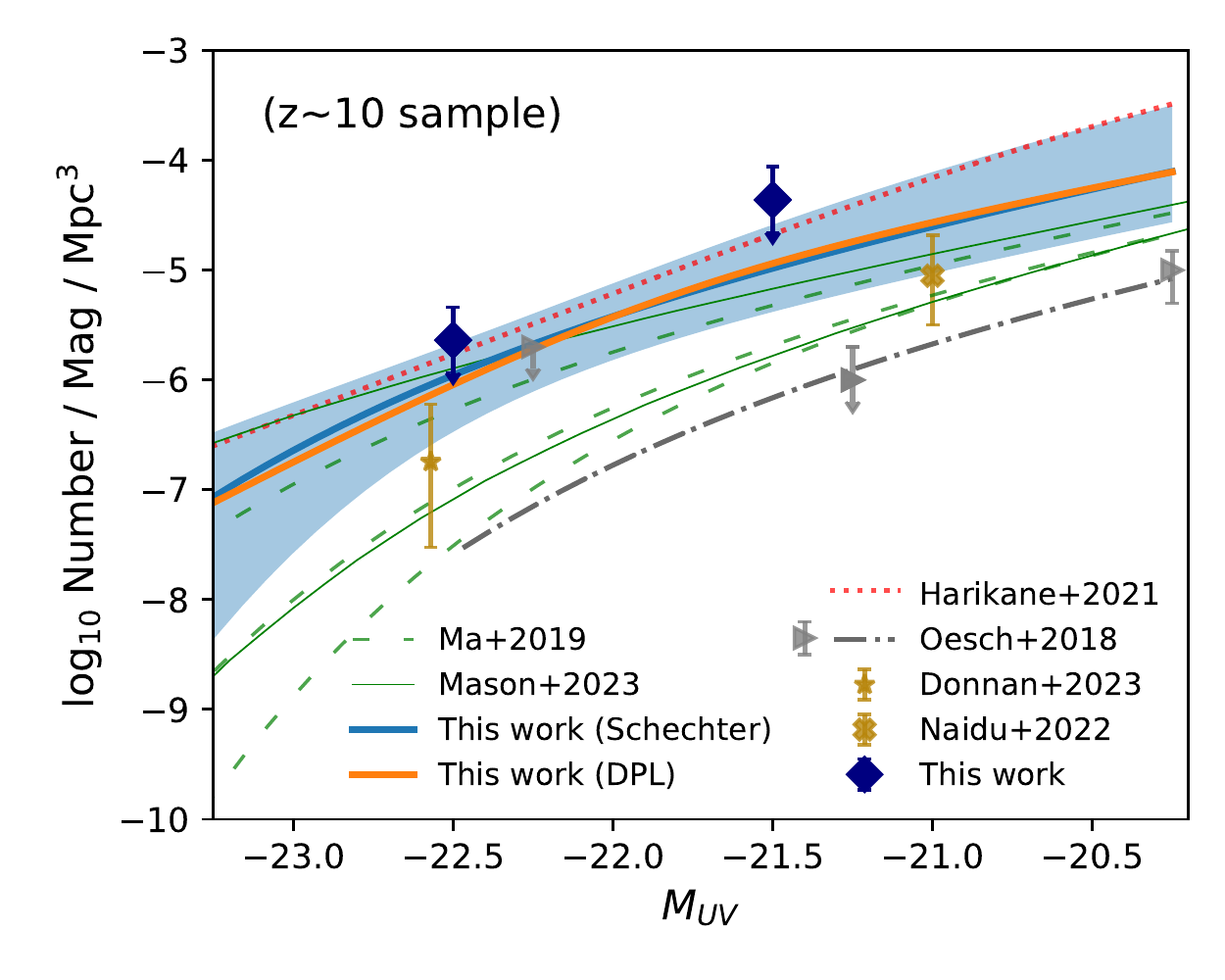}
    \caption{Similar to Figure \ref{fig:derived_UVLFs_z8} but for UVLFs at $z=10$. The gold cross shows the results from early JWST observations from \citet{Naidu2022}.}
    \label{fig:derived_UVLFs_z10}
\end{figure}

The best-fit parameters are the medians of the burned-in parameters stacked together from all the MCMC steps and Monte Carlo iteration. We list them in Table \ref{tab:best-fit_parameters} and show the best-fit Shechter functions (blue lines) and the best-fit DPL functions (orange lines) in Figure \ref{fig:derived_UVLFs_z8}-\ref{fig:derived_UVLFs_z10}. The blue shades encompass $1\sigma$ uncertainties of the best-fit Schechter functions. The navy data points are nominal luminosity functions derived with the method similar to those in \citet{Rojas-Ruiz2020} (see Appendix \ref{appendix:other_methods}).


\begin{table}[]
\begin{tabular}{ccccc}
\hline
$z$ & $\phi$                & $M^*$ & $\alpha$ & $\beta$ \\
    & ($10^{-4}$/mag/Mpc$^3$) & (mag) &  (fixed) &                      \\ \hline\\
8 & $1.40^{+4.76}_{-1.08}$ &
          $-22.30^{+0.75}_{-0.92}$ & -2.23 &  \\[3pt]
8 & $1.29^{+7.58}_{-0.91}$ &
          $-22.15^{+1.02}_{-0.75}$ & -2.23 & 
          $-4.74^{+1.41}_{-1.83}$ \\[3pt]
9 & $1.71^{+20.33}_{-1.57}$ &
          $-21.87^{+0.92}_{-1.28}$ & -2.33 &  \\[3pt]
9 & $1.31^{+19.64}_{-1.18}$ &
          $-21.87^{+1.07}_{-1.20}$ & -2.33 & 
          $-4.93^{+1.60}_{-1.71}$ \\[3pt]
10 & $1.16^{+17.88}_{-1.08}$ &
          $-21.98^{+0.94}_{-1.22}$ & -2.38 &  \\[3pt]
10 & $1.10^{+20.15}_{-1.02}$ &
          $-21.88^{+1.13}_{-1.24}$ & -2.38 & 
          $-4.84^{+1.58}_{-1.75}$ \\[3pt] \hline
\end{tabular}
\caption{\label{tab:best-fit_parameters} The best-fit parameters for the Schechter functions (upper rows) and the best-fit parameters for the DPL functions (lower rows) at each redshift. The faint-end slopes ($\alpha$) are fixed to the values derived by \citet{Bouwens2021}.}
\end{table}

\section{Discussion}
\label{sec:discussion}
Our best-fit luminosity functions at $z\sim8-10$ for the relatively faint bin of the derived UVLF at $z\sim8$ with \MUV$\gtrsim-22$ agree well with previous studies. 

In contrast, our results at the brightest bins are larger than both DPL and Schechter functions derived from many earlier HST and ground-based observations, e.g. \citet{Bouwens2021}, \citet{Bowler2020} and \citet{Varadaraj2023}, but less than that derived by \citet{Rojas-Ruiz2020}. \citet{Bouwens2021} and \citet{Oesch2018} (also \citet{Finkelstein2015}) derived their luminosity functions from deep HST imaging such as the Hubble Ultra Deep Field, the Hubble Frontier Fields parallel fields, and the CANDELS fields. Though they include $\sim200$ arcmin$^2$ of HST parallel data, the majority of their area is from the CANDELS fields, whose depths range from similar to 1 mag deeper than our SuperBoRG dataset. Most of their observations come from a few contiguous areas, while all of ours are spread across over 200 independent sightlines. Similarly, the observational data in \citet{Stefanon2019} and \citet{Bowler2020} come from ground-based imaging over one to two contiguous areas (including the COSMOS field). Their depths in the J band are roughly $\sim0.5$ to 2 mag shallower than those of SuperBoRG. The grey circles in Figure \ref{fig:derived_UVLFs_z8} are from \citet{Rojas-Ruiz2020}. That work was based on public archival HST data from the BoRG dataset, which covered about a third of the SuperBoRG data we used here. It was derived independently from the BoRG team using different reduction pipelines, selection methods, and completeness simulations. Our results agree with their $z\sim8$ UVLF at \MUV$\gtrsim-22$ but shows smaller number density at the brightest bin \MUV$\sim-23$ with at $\sim1\sigma$ confidence.

Early results from the James Webb Space Telescope (JWST) observations suggest that there are more bright galaxies at very high redshift ($z\gtrsim12$) than predictions from theoretical models and extrapolations of lower-$z$ observations in legacy fields \citep{Castellano2022,Mason2023, Wilkins2023, Harikane2023}. As for UVLFs at $z=8-10$, \citet{Donnan2023} used 45 arcmin$^2$ of JWST/NIRCam observations from Early Release Observations (ERO) and Early Release Science (ERS) programmes. Their UVLFs at $z=8$ and $z=9$ (gold stars in Figure \ref{fig:derived_UVLFs_z8} and Figure \ref{fig:derived_UVLFs_z9}) are consistent with results from non-BoRG data sets. However, \citet{Harikane2023}, who used later versions and larger areas that total 90 arcmin$^2$ of ERO and ERS observations, derived UVLFs at $z=9$ with much higher number density that agrees well with our results (gold pentagons in Figure \ref{fig:derived_UVLFs_z9}). Interestingly, their uncertainties are larger than those estimated by \citet{Donnan2023} despite coming from a larger area of the same programs. 
At $z=10$, the JWST-based determinations from \citet{Donnan2023} and \citet{Naidu2022} show large uncertainties resulting in an agreement with both our UVLF and the UVLF determined by \citet{Oesch2018}.

One question worth checking is whether HST parallel observations are subjected to systematic bias (i.e. instrumental systematic, typical relative depths in the filters) or whether the difference has a physical origin (in particular cosmic variance). The former would mean that our sample still has contamination that has not been accounted for through our analysis. One possible explanation could be the limited number of blue filters typically available in parallel surveys. About two-thirds of our final candidates are in the fields observed with only a single filter bluer than the dropout one (specifically a very wide F350LP). In those conditions, it is the only band used to determine the non-detection of the candidate below the Lyman limit criterion. Given the limited total time of each parallel observation, a long-pass filter F350LP, which is sensitive from 3500 to 8000\AA, is advantageous for detecting light coming from interlopers over a range of wavelengths. Indirectly, some support for this possibility comes from the comparison to the UVLF at $z\sim 8$ from early BoRG observations, which were based on F098M as a dropout filter and F606W as a blue filter \citep{Trenti2011, Bradley2012, Schmidt2014}. Nonetheless, having a single long-pass filter may raise a concern that low-z galaxies could leak into our sample -- a case has already been illustrated by early JWST verifications of some especially luminous and high redshift candidates from deep fields with a larger assortment of filters \citep{ArrabalHaro23}. However, as shown in Section \ref{sec:potential_contamination}, we have already taken this into account by estimating field-by-field numbers of possible low-z interlopers using a recent mock catalogue that was based on the currently known properties of low-z galaxies. A possible exception that would make our contamination estimation too low is that there is a large number of hidden dusty star-forming galaxies, as suggested by a numerical model \citep{Long2022} that is yet to be confirmed by observations. And that the hidden population only leaks into the sample using our observation setups.


Next, we investigate whether cosmic variance can explain the discrepancies. Brighter galaxies cluster more strongly and more subjected to cosmic variance \citep{TrentiStiavelli2008, Robertson2014, Park2017}. Observational determination of galaxy clustering through the two-point correlation function is currently limited to $z\lesssim7$ due to a small number of known galaxies at higher redshifts \citep{Barone-Nugent2014, Harikane2021}. Hence, we resort to findings from simulations. \citet{Bhowmick2020} used \texttt{BLUETIDES}, which is the largest cosmological hydrodynamic simulation for galaxy formation in terms of volume, to estimate the effect of cosmic variance on $z>7$ galaxy surveys. They found that for the survey area of 100 arcmin$^2$ to 1 deg$^2$, the uncertainties in the number count of galaxies brighter than \MUV\ $\sim-22$ mag due to cosmic variance is to-the-order-of-magnitude comparable to the uncertainties in the number count due to Poisson variance. Given that the area of the SuperBoRG data set is in this range, cosmic variance can thus affect the number of candidates at \MUV$\leq-22$ mag by at most a factor of a few (half a dex), 
which is just enough to bring the lower bound of our $z=8$ best-fit function to be consistent within $1\sigma$ with the results from targeted-area observations. However, the effect of cosmic variance in our data should already be minimal by the design of the parallel observations and smaller than the predicted half a dex that was based on a contiguous area. Given that the observed area in \citet{Bowler2020} is $\sim$ 6 deg$^2$, it is also unlikely that cosmic variance will change their number counts significantly, except perhaps the brightest magnitude bins \MUV$\sim-23$. Therefore, based on the cosmic variance estimated from the \texttt{BLUETIDES} simulations, cosmic variance is unlikely, but possible in the extreme case, to explain the entire amount of the discrepancies in the observed UVLFs at $z\sim8-10$. 

If the large number density at the bright end is not due to cosmic variance, another possible explanation is a contribution from AGNs to our UVLFs. Although our samples were selected with some stellarity cut, it was mainly to mitigate the contamination from galactic stars. At faint magnitude ($m>24$), the stellarity parameter cut of \texttt{CLASS\_STAR}$<0.95$ may still allow quasars and compact star-forming galaxies to be included in the sample \citep{Morishita2021, Ishikawa2022}. \citet{Harikane2021} derived rest-UV luminosity functions of Lyman-break dropout sources (which include both AGNs and galaxies) at $z\sim4-7$ from the Subaru/Hyper Suprime-Cam survey. They found that both DPL+DPL and DPL+Schechter functions fit the shape of their luminosity functions well. Using spectroscopic observations to determine the AGN fractions in their data set, they concluded that the dropout luminosity function is a superposition of an AGN luminosity function and a galaxy luminosity function. AGNs dominate the bright end, while galaxies, whose intrinsic shape is of DPL or Schechter form, dominate the faint end. The AGNs start to contribute to the dropout luminosity function at \MUV$\sim-21.5$ and entirely dominate the population at \MUV$\sim-24$ at all redshifts.
Although \citet{Harikane2021} found that the number densities of quasars rapidly decreased from $z\sim3$ to $z\sim6$ and concluded that bright sources at $z\sim8-10$ are likely dominated by galaxies, their number density of bright sources in the AGN-dominated area at $z\sim6$ actually shows very similar density as those at $z\sim7$. This is demonstrated in dotted yellow and dotted red lines in Figure \ref{fig:derived_UVLFs_z8}, where at the bright end, the number density of dropouts at $z=7$ (red dotted line) is very similar or even larger than that at $z=6$ (yellow dotted line). Our result further extends this finding to $z\sim8$. The bright end of our derived $z\sim8$ UVLF is consistent with their $z\sim6-7$ UVLFs. 

We nonetheless note that \citet{Varadaraj2023} found that some of the brightest $z=7$ candidates in \citet{Harikane2021} are better fit with cool dwarf templates than galaxy templates when redder bands are included. If these candidates are later confirmed to be contaminants, the $z=7$ UVLF in \citet{Harikane2021} would be lower and that the high density of bright $z=8$ galaxies found here would be even more perplexing (or is subjected to higher contamination than expected).
Another supporting evidence that AGNs could contribute to the large number of densities at the bright end of UVLF comes from the recently discovered AGN in a $z=8.7$ galaxy with $M_\textrm{UV}\approx-22$ mag \citep{Larson2023}. They found that the AGN contributes approximately $30\%$ of the total rest-UV continuum emission from the galaxy. This suggests that even for sources just above the characteristic $L_*$ luminosity (i.e. $M_\textrm{UV}\approx-22$), AGNs could still partially contribute to the UVLFs by enhancing the UV fluxes. Therefore, it is possible that AGNs could contribute to the large number density at the bright end found in this work.

Compared with theoretical predictions, our best-fit UVLFs agree well with models with moderate to low dust attenuation. Dust attenuation
has been suggested as a main contributor to the variation at the bright end of UVLFs  \citep[e.g.][]{Tacchella2018, Yung2019}. Semi-analytical and hydrodynamical simulations generally agree that dust fractions must drop at high redshifts $z\gtrsim9$. There is not enough cosmic time for most low- to intermediate-mass stars to reach the asymptotic giant branch stage, which is the main dust production site in the local universe \citep{Dwek2014, Vogelsberger2020}. \citet{Ma2019} applied a dust radiative transfer code to the FIRE-2 simulations to study the effect of dust-to-metal ratios ($f_\textrm{dust}$) on both rest-frame IR and UV properties. Their predicted UVLFs at $z\sim8-10$ at different $f_\textrm{dust}$ values are shown in the three green loosely-dashed lines in Figure \ref{fig:derived_UVLFs_z8} -- \ref{fig:derived_UVLFs_z10}. From top to bottom in each figure, the lines represent $f_\textrm{dust} = 0, 0.4,$ and 0.8, respectively. Separately, \citet{Mason2023} used a semi-empirical model where star formation is set by halo mass. Their predicted UVLFs with and without dust attenuation are shown as lower and upper green solid lines, respectively, in Figure \ref{fig:derived_UVLFs_z8} -- \ref{fig:derived_UVLFs_z10}. Compared with \citet{Ma2018}, the bright ends of our UVLFs are most consistent with the models with $f_\textrm{dust}$ smaller than or equal to 0.4, which is the canonical dust-to-metal ratio in the local Universe \citep{Dwek1998}. Interestingly, our UVLF at $z=10$ seems to be more dust poor than those at lower redshifts as it agrees more with the $f_\textrm{dust}=0$ case. Comparing with \citet{Mason2023}, we find a similar conclusion. The best-fit UVLFs generally fall in between their with- and without-dust UVLFs. Since \citet{Mason2023} included the effect of dust via an empirical dust attenuation at $z\sim3$, their with-dust prediction may be based on higher attenuation than true $z\gtrsim8$ galaxies.

If dust is most responsible for the number density of the UVLFs at the bright end, our observation supports the conclusion that $z\gtrsim8$ galaxies are dust poor. This is in line with other observational results. \citet{Behrens2018} used a radiative transfer code to reproduce an observation of an observed (seemingly) dusty galaxy at $z=8.4$ \citep{Laporte2017}. They found that the dust-to-metal ratio required is only 0.08, suggesting that early dust formation is inefficient. \citet{Leethochawalit2023} also found that JWST-selected $z\sim8$ galaxies have dust extinction that is a few times lower than typical values at $z=2-3$. For JWST-selected galaxies, \citet{Furtak2023} suggested that $z\gtrsim9$ galaxies have very blue UV-slopes and extremely low dust attenuation with $A_V \leq 0.02$. Similar results are found by \citet{Santini2023} and, at lower redshift, \citet{Nanayakkara2023}.

Regardless, not all observational results suggest galaxies have lower dust content at high redshifts. Stellar population studies of bright $z>8$ galaxies have tentatively indicated that these galaxies have SFHs that extend beyond a redshift $z=10$. The extended SFH can allow galaxies more time to build up their dust content \citep{RB20, Laporte2021, Tacchella2021}. Furthermore, the UV slopes of $z\gtrsim8$ galaxies seem to not differ from those at lower redshifts $z=4-8$, which may suggest that the two populations have similar dust attenuation \citep{Roberts-Borsani2021, Tacchella2021}. 

Other possible factors that can significantly alter the UVLFs at the bright end include starburst timescale and initial mass functions (IMFs). \citet{Lacey2016} used a semi-analytical model of galaxy formation to examine how different parameters for physical processes affect the behaviour of UVLFs at $0<z<6$. Among the parameters they considered, two can change the bright end of the luminosity function by more than two dex: starburst timescale and IMF of starbursts. When a shorter timescale of starbursts is allowed, e.g. 0.05 Gyr as opposed to 0.5 Gyr, the number density at \MUV$=-21$ mag at $z=6$ increases by more than two dex. Besides, if the IMF of starbursts is changed to be Kennicutt-like (i.e. having relatively smaller numbers of massive stars), the number density at \MUV$=-22$ mag can be reduced by as much as $\sim2$ dex.

In terms of the parameters and the functional forms that describe the UVLFs, Figure \ref{fig:derived_UVLFs_z8}-\ref{fig:derived_UVLFs_z10} shows the best-fit Schechter functions and the best-fit DPL functions in blue and orange. At all redshifts, the best-fit Schechter and best-fit DPL functions are almost identical and can describe our observed sample well. This contrasts with the finding by\citet{Bowler2020} that the DPL function is a better fit than a Schechter function. The reason for this is that our derived characteristic magnitude $M^*$ is about two magnitudes brighter than that derived by \citet{Bowler2020} and \citet{Bouwens2021}, which delays the onset of the exponential drop off of the Schechter function. We nonetheless note that our best-fit functions are only limited to the relatively bright end ($M_\textrm{UV}\lesssim20.5$). Such limited range may cause us not to be able to differentiate between the two functional forms. Regardless, when compared to the predictions in \citet{Ma2019}, the derived characteristic magnitudes $M^*$ are again consistent with their best-fit values of the low-dust models ($f_\textrm{dust}=0$ to 0.4). 

\section{Conclusions}

In this work, we derived UV luminosity functions at $z=8-10$ based on the largest compilation of HST (pure-)parallel observations from the SuperBoRG project. Depending on filter requirements for dropout selection at each redshift, the search area we considered ranges from 840 arcmin$^2$ to over 1300 arcmin$^2$. We identify candidates via a dropout selection (colour and signal-to-noise cut) first and further refine the sample by visual inspections and a photometric redshift selection. We carefully checked for cool dwarf stellar contamination by comparing the best-fit dwarf spectra to the best-fit galaxy spectra and estimating the probability of being dwarfs based on additional assumptions on sky densities of dwarfs and LBGs. We further estimate the chance of having low-z galaxy interlopers in the field of each candidate. The $1\sigma$ range of the final probabilities that the candidates are true high-z galaxies is 58\% to 95\%. Source recovery simulations show that sample completeness is $\sim50-70\%$ at the bright end and starts to drop off at \MUV$\sim-21.5$. The best-fit UVLFs are shown in Figure \ref{fig:derived_UVLFs_z8}--\ref{fig:derived_UVLFs_z10}. Our main findings are as follows:

\begin{enumerate}
    \item The derived UVLFs at the bright end are up to $\sim1$ dex higher than the UVLFs previously derived based on observations in legacy fields (from the ground or space) for $L>>L_*$, but smaller than other independent analysis of part of the SuperBoRG dataset. At magnitudes fainter than -22 mag (or $L\lesssim L_*$), the derived UVLFs are with all literature determinations. The difference at the very bright end of the UVLFS could still be attributed primarily to cosmic variance, but this is unlikely. It also could be partially due to the lack of multiple blue filters in the HST parallel observations if there is an unidentified class of contaminants based on rare intermediate redshift galaxies with a very high dust content.
    
    \item We proposed two explanations for the physical origin of the excess at the bright end. First is the contribution from AGNs, where our UVLF determination is consistent with the extrapolation of recent measurements of the AGN LF, e.g. by \citet{Harikane2021}. The derived UVLF at $z\sim8$ is consistent with the lack of evolution from $z=6-7$ at the bright end once AGN contributions are included in the UVLF determination. Second is that the galaxies at $z\geq8$ are dust poor. The derived UVLFs at all redshifts are consistent with theoretical predictions based on an evolving dust-to-metal ratio at $z\gtrsim 8$ so that the galaxies we observe have lower dust and/or dust-to-metal content than their counterparts at $z\lesssim 7$. 
    \item For our UVLFs, the best-fit Schechter functions are statistically indistinguishable from best-fit DPL functions at all redshifts and both describe our data well. 

\end{enumerate}

To understand the origin of the large number densities of the brightest galaxies in our observed UVLFs, spectroscopic follow-up is critically required. Early JWST NIRCam and NIRSPEC observations indicate that there may be a large population of massive systems in the early universe \citep{Boyett2023}. Further spectroscopy will unveil whether our candidates are AGN-dominated, low-dust galaxies or unforeseen contaminants, providing valuable insight into early galaxy formation and evolution irrespective of the outcome of such follow-ups.

\section{Acknowledgements}

This research was supported by the Australian Research Council Centre of Excellence for All Sky Astrophysics in 3 Dimensions (ASTRO 3D) through project number CE170100013. We acknowledge partial support from NASA through grant JWST-ERS-1342. Lastly, this research used Spartan, a High-Performance Computing system operated by Research Computing Services at The University of Melbourne \citep{Lafayette2016}.

\section{Data Availability}
The data underlying this article will be shared on reasonable request to the corresponding author.



\bibliographystyle{mnras}
\bibliography{references}




\appendix
\section{Other UVLFs derivation methods}
\label{appendix:other_methods}
This section derives UVLFs with two other common methods in the literature. As shown in \citet{Leethochawalit2022}, different methods may yield systematically different UVLFs depending on the intrinsic scatter of the UV luminosities. Our fiducial method in Section \ref{sec:UVLF} was shown to be insensitive to flux scatter. However, it may be subject to stochasticity at the extremely bright end. Therefore, we cross-check the results with two other approaches. 

\begin{figure}
    \centering
    \includegraphics[width=0.85\textwidth]{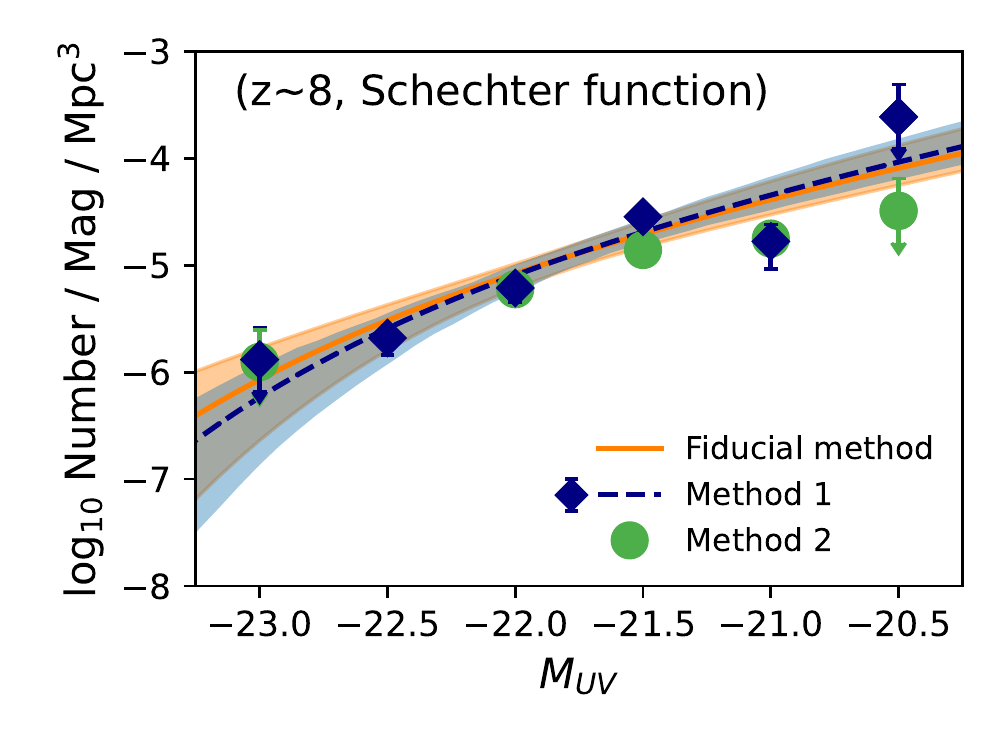}\\
    \includegraphics[width=0.85\textwidth]{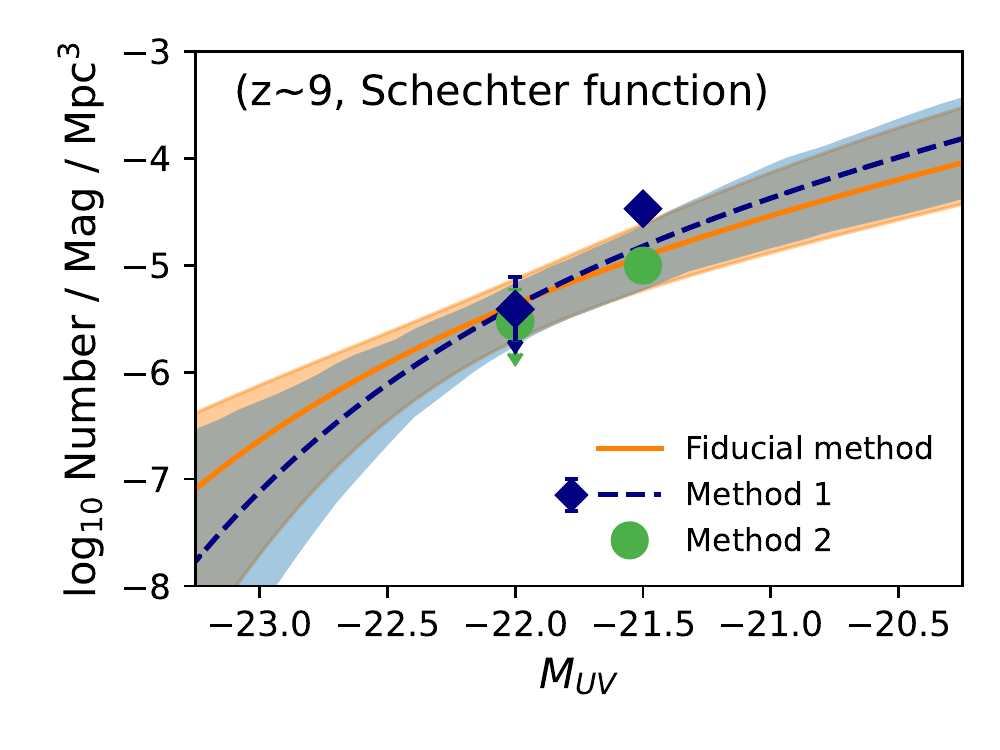}\\
    \includegraphics[width=0.85\textwidth]{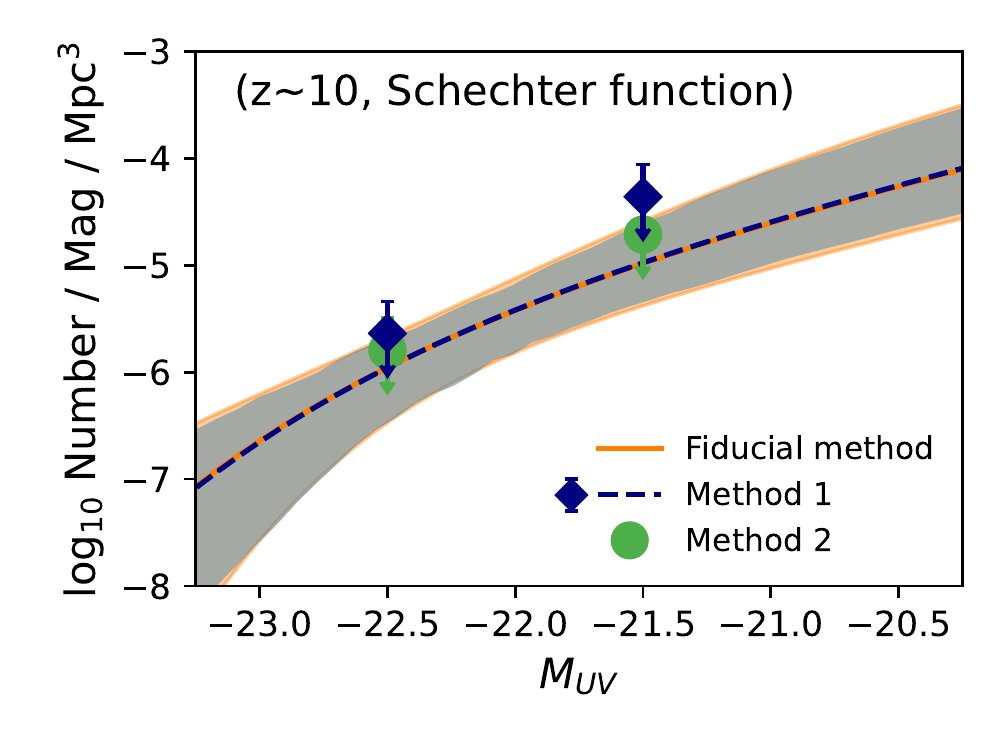}
    \caption{Comparison of UVLFs derived by three different methods in the literature. All best-fit parameters (lines and shaded regions) are of Schechter function.}
    \label{fig:all_methods_UVLF}
\end{figure}

The first method we consider here is similar to what was done in \citet{Finkelstein2015} and \citet{Rojas-Ruiz2020} (grey circles in Figure \ref{fig:derived_UVLFs_z8}). It uses completeness as a function of intrinsic magnitude and redshift $P(M_\textrm{in},z)$, which is defined as the fraction of recovered galaxies at each input magnitude $M_\textrm{in}$ and redshift $z$ bin. An example of this completeness is shown in Figure \ref{fig:completeness_Minput_example}. For this method, the effective volume is 
\begin{equation}
    V_\textrm{eff}(M_\textrm{in}) = \int \frac{dV}{dz}P(M_\textrm{in},z) dz.
\end{equation}
where $V$ is the comoving volume at redshift $z$ associated with the image. With the effective volume summed over all the fields in the survey, the nominal luminosity function can be calculated as
\begin{equation}
    \phi(M_\textrm{in}) = N^\textrm{obs}(M_\textrm{in})/V_\textrm{eff}(M_\textrm{in})/dM.
\end{equation}
Here, $N^\textrm{obs}(M_\textrm{in})$ is the number of candidates with intrinsic UV magnitudes in the $M_\textrm{in}$ bin. To obtain the intrinsic UV magnitude of each candidate, we correct the measured UV magnitude with the mean flux bias found in the completeness simulation for its field, i.e. those that constitute the box plots in the lower panels of Figure \ref{fig:fluxscatter}. We assume that the number of observed galaxies follows the Poisson distribution and that the uncertainty of $N^\textrm{obs}(M_\textrm{in})$ is the square root of the number. We plot the resulting nominal luminosity functions as navy diamonds in Figure \ref{fig:derived_UVLFs_z8}-\ref{fig:derived_UVLFs_z10} and Figure \ref{fig:all_methods_UVLF}. To obtain the best-fit Schechter and DPL functions using this definition of completeness, we compare the modelled number of observed galaxies 
\begin{equation}
    N^\textrm{model}(M_\textrm{in}) =
    \phi(M_\textrm{in})\cdot V_\textrm{eff}(M_\textrm{in}) \cdot dM
\end{equation}
to the actual $N^\textrm{obs}(M_\textrm{in})$. Similar to what was done in Section \ref{sec:UVLF}, we use the MCMC to find the best-fit parameters by assuming a Poissonian log-likelihood distribution  (Equation \ref{eq:poisson}). We also resample the candidates and run the MCMC 100 times for each low contamination-by-dwarfs and high contamination-by-dwarfs scenario. We then combine all the burned-in parameters from these MCMC chains to find the final best-fit model parameters. The best-fit Schechter functions from this method are shown as dashed blue lines in Figure \ref{fig:all_methods_UVLF}.

The second method uses the effective volumes at the recovered or measured UV magnitudes (Equation \ref{eq:veff_mrecov}, Figure \ref{fig:effective_volume}). Examples of past studies that used this method include \citet{Calvi2016, Morishita2018}. In this case, the nominal luminosity function is
\begin{equation}
    \phi(M_\textrm{in}=M_\textrm{recov}) = N^\textrm{obs}(M_\textrm{recov})/V_\textrm{eff}(M_\textrm{recov})/dM.
\end{equation}
For this method, the translation between recovered and intrinsic magnitudes is encoded in the $V_\textrm{eff}(M_\textrm{recov})$ directly. So there is no need to make corrections to the measured UV magnitudes of the candidates. Here, we also assume that the uncertainty of the $N^\textrm{obs}(M_\textrm{recov}$ is the square root of the number.

The nominal and best-fit Schechter functions from the three methods considered are shown in Figure \ref{fig:all_methods_UVLF}. The best-fit relations are consistent with each other. The only minor caveat is that the nominal UVLFs derived by method 2 seem slightly flatter than method one on the fainter side. 


\bsp	
\label{lastpage}
\end{document}